\documentclass[final,5p,times,english,twocolumn]{elsarticle}

\usepackage{amssymb}
\usepackage{tensor}

\usepackage{amsmath}
\usepackage{xcolor}
\usepackage{subcaption}
\usepackage{dsfont}
\usepackage{mhchem}
\usepackage{background}
\usepackage{todonotes}
\usepackage{mathtools}
\journal{arXiv}

\usepackage{hyperref}
\hypersetup{
     colorlinks=true,
      linkcolor=blue,
}

\renewcommand\bold\mathbf

\backgroundsetup{contents=}
\begin{document}

\begin{frontmatter}




\title{Properties of non-cryogenic DTs and their relevance for fusion}


\author{Hartmut Ruhl$^1$, Christian Bild$^1$, Ondrej Pego Jaura$^1$, Matthias Lienert$^1$,
Markus N\"oth$^1$, Rafael Ramis Abril$^2$, and Georg Korn$^1$} 

\address{$^1$Marvel Fusion, Theresienh\"ohe 12, 80339 Munich,
  Germany\\
$^2$E.T.S.I. Aeronáutica y del Espacio, Universidad Politécnica de Madrid, P. Cardenal Cisneros 3, Madrid 28040, Spain}

\begin{abstract}
In inertial confinement fusion, pure deuterium-tritium (DT) is usually
used as a fusion fuel. In their paper \cite{gus2011effect}, Guskov et
al. instead propose using low-Z compounds that contain DT and are
non-cryogenic at room temperature. They suggest that these fuels (here
called non-cryogenic DTs) can be ignited for $\rho_{\ce{DT}} R  \geq
0.35 \,\ce{g} \, \ce{cm^{-2}}$ and $kT_{\ce{e}} \geq 14 \, \ce{keV}$,
i.e., parameters which are more stringent but still in the
same order of magnitude as those for DT. In deriving these results the
authors in \cite{gus2011effect} assume that ionic and electronic
temperatures are equal and consider only electronic stopping
power. Here, we show that at temperatures greater than 10 keV, ionic
stopping power is not negligible compared to the electronic one. We
demonstrate that this necessarily leads to higher ionic than
electronic temperatures. Both factors facilitate ignition compared to
the model used in \cite{gus2011effect} showing that non-cryogenic DT
compounds are more versatile than previously known. In addition, we
find that heavy beryllium borohydride ignites more easily than heavy
beryllium hydride, the best-performing fuel found by Guskov et al.
Our results are based on an analytical model that incorporates a
detailed stopping power analysis, as well as on numerical simulations
using an improved version of the community hydro code MULTI-IFE.
Alleviating the constraints and costs of cryogenic technology and the fact that
non-cryogenic DT fuels are solids at room temperature open up new
design options for fusion targets with $Q>100$ and thus contribute to the larger
goal of making inertial fusion energy an economically viable source of
clean energy. In addition, the discussion presented here generalizes
the analysis of fuels for energy production.
\end{abstract}

\begin{keyword}
nuclear fusion, ignition of non-cryogenic DT, high fusion gain



\end{keyword}

\end{frontmatter}


\tableofcontents

\section{Introduction}
In search of fusion fuels which are solids at room temperature, we
have recently started to investigate a class of non-cryogenic chemical
compounds in \cite{ruhlkornarXiv0,ruhlkornarXiv1,ruhlkornarXiv2,ruhlkornarXiv3}.
They consist of elements with charge numbers $Z$ of 1 to 5 which bind
hydrogen in solid form. As deuterium and tritium have chemically almost
identical properties as hydrogen, we assume that it is possible to
replace the hydrogen atoms by deuterium and tritium in equal
ratios. We call the resulting compounds \emph{non-cryogenic DTs}.

In \cite{ruhlkornarXiv0,ruhlkornarXiv1,ruhlkornarXiv2,
ruhlkornarXiv3} the focus has been on non-igniting inertial fusion
energy (IFE) targets with gain $G \approx 1$ without fuel
pre-compression preheated with only a few $\ce{MJ}$ to
highlight the interesting properties of non-cryogenic DTs.
However, as is widely accepted, high gain $G\gg 1$ in IFE on a
practical level rests on at least two points: I) The fuels must be
capable of igniting and II) fuel compression is needed to
reduce the fuel mass to a level acceptable for ignition and energy
production. The present paper focuses on the first of these
aspects and studies several practical compounds in greater
depth beyond the more academic fuels initially addressed in
\cite{ruhlkornarXiv0,ruhlkornarXiv1,ruhlkornarXiv2, ruhlkornarXiv3}.

Non-cryogenic DTs are compounds with an inactive low-Z and an active DT
component. Their total mass density is given by $\rho=\rho_{\ce{Z}} + \rho_{\ce{DT}}$,
where $\rho_{\ce{Z}}$ denotes the low-Z and $\rho_{\ce{DT}}$
the $\ce{DT}$ mass density. Table \ref{tab:noncryogenicdt} provides a
non-exhaustive list of non-cryogenic DT compounds which, as we show,
can ignite, reach high fusion gain and thus serve as potential fusion
fuels. They are chosen for their high mass density of $\ce{DT}$
relative to the mass density of the low-$Z$ elements. In fact, their
$\ce{DT}$ density is higher than the one of $\ce{DT}$ ice.

The fuel compounds we consider should be distinguished from so-called
contaminated fuels, see e.g. \cite{pasley2011,
khatami2020}. In our case, the presence of $Z>1$ ions is essential for the
properties of the fuel, combining the downside of increased radiation 
losses with significant upsides like the fuel being a solid material
at room temperature and having increased stopping power, whereas for
contaminated fuels the presence of $Z>1$ is usually unwanted.

Non-cryogenic chemical compounds with inactive impurities
and active $\ce{DT}$ for fusion applications have previously been
addressed by Guskov et
al. \cite{gus2011effect,gus2013compression,gus2015fast,gus2016influence}.
The authors call these compounds low-Z fuels and find that they are
capable of igniting and of producing gain (see \cite[table
2]{gus2011effect}) assuming equal electron and ion
temperatures $kT_{\ce{e}} = kT_{\ce{i}} = kT$. The authors assume that
stopping of the fusion $\alpha$-particles is only due to electrons.
However, as we show here, ionic fuel temperatures can be
considerably higher than the electronic one, implying $kT_{\ce{i}} >
kT_{\ce{e}}$. Moreover, ionic $\alpha$-particle stopping becomes
relevant in non-cryogenic DT at its elevated ignition temperatures.
Including these effects, we improve the versatility of low-Z fuels
beyond the predictions made in \cite{gus2011effect,gus2015fast}
since the effective fusion power grows relative to the power loss by
radiation for $kT_{\ce{i}} > kT_{\ce{e}}$. What is more, the chemical
compound that performs best in this study, heavy beryllium
borohydride, needs about \(10\%\) lower confinement parameter
\(\rho_{\mathrm{DT}} R\) to be ignited compared to heavy beryllium
hydride, the compound advocated for in
\cite{gus2011effect,gus2015fast}. It also requires about \(10\%\)
lower ignition energy due to a lower number of electrons and ions
per pair of deuterium and tritium ions.

The paper is structured as follows. In sec. \ref{sec:ignition} we
introduce our analytical model for ignition for the ideal case of a
pre-compressed static plasma and neglecting fuel consumption. This
includes a novel analysis of ionic and electronic stopping
powers for $\alpha$-particles and neutronic energy deposition in the
fuel. Importantly, these simplifications allow for a temperature flow
analysis in a static geometry, enabling us to identify a
quasi-equilibrium of ion and electron temperatures with $kT_i >
kT_e$. Using this analysis, we identify the minimal confinement
parameters $\rho_{\ce{DT}} \, R$ for ignition as well as the corresponding
ignition temperatures for the selected non-cryogenic DT
compounds. Moreover, we contrast the energy deposition to ions and
electrons for non-cryogenic DT fuels with that for DT. In
sec. \ref{sec:numerical}, we compare this analytical model with
numerical simulations using an improved version of the MULTI-IFE hydro
code for a pre-compressed, static plasma considering fuel consumption.
In sec. \ref{sec:rescale} we make the MULTI-IFE predictions
plausible with the help of simple analytical considerations.
We conclude in sec. \ref{sec:summary}. While the main text is kept
brief, theoretical derivations and technical details are included in dedicated
appendices.

\begin{table}[]
    \small
    \centering
    \begin{tabular}{c|c|c|c}
      name & composition & \( \rho \) 
      & \( \rho_{\ce{DT}}\)\\ 
      \hline 
      deuterium-tritium & DT & 0.225 & 0.225\\
      beryllium borohydride & $\ce{Be}[\ce{BD}_2\ce{T}_2]_2$ & 0.790 & 0.310\\
      beryllium hydride & BeDT & 0.961 & 0.343\\
      lithium borohydride & $\ce{LiBD}_2\ce{T}_2$ & 0.848 & 0.303\\
      \hline 
      \end{tabular}
\caption{Non-cryogenic DT compounds, their total mass
densities $\rho=\rho_{\ce{Z}}+\rho_{\ce{DT}}$ and their $\ce{DT}$ mass
densities $\rho_{\ce{DT}}$ bound in the compound. The mass densities
have been calculated from chemical tables
\cite{atzeni2004physics,zuttel2003hydrogen,smith1988crystal,marynick1972crystal}
by exchanging hydrogen by deuterium and tritium. The mass densities 
of $\ce{BeDT}$ have been obtained using the crystalline state
density of $\ce{BeH_2}$. All densities are in units of $\ce{g}\, \ce{cm^{-3}}$.}
\label{tab:noncryogenicdt}
\end{table}

\section{Analytical model for ignition} \label{sec:ignition}

\paragraph{Overview of the model}
Ignition is the rise of temperatures due to the fusion power deposited to the plasma
exceeding the combined power losses. Here we consider the power balance for a
spherical fusion plasma with confinement parameter $\rho_{\ce{DT}} R$ as can be reached in the stagnation phase of fuel pre-compression. We include radiation losses
by bremsstrahlung $Q_{\text{rad}}$ given by \eqref{loss:rad}, losses due to escaping
$\alpha$-particles at the surface by making use of the deposition powers $Q_{\ce{fus \rightarrow i}}$ given by \eqref{fusionpower}
and $Q_{\ce{fus \rightarrow e}}$ given by \eqref{fusionpower}, electronic heat
loss at the surface $Q_{\ce{e}}$ given by \eqref{loss:conduction}, and
heat transfer $Q_{\ce{bc}}$ between the different particle species in the plasma
given by \eqref{eq:Q_bc} (see \ref{sec:gainloss}). The deposition powers to
ions $Q_{\mathrm{fus}\rightarrow i}$ and electrons
$Q_{\mathrm{fus}\rightarrow e}$ are calculated in detail according to
a novel stopping power model for alpha particles and neutrons, making use of
the Euler equations \eqref{euler1}--\eqref{euler3}
for spherical geometry (see \ref{sec:analyticalmodel}). We neglect hydro motion as well as fuel depletion. Importantly, these simplifications allow for a
temperature flow analysis in an otherwise static situation. To justify the application of this analysis to a fusion
scheme, one needs to assume that the stagnation phase is sufficiently
long for ignition to happen and that fuel depletion is not too
severe. To include fuel depletion and to cross-check the
model, we perform one-dimensional MULTI-IFE simulations neglecting hydro motion in sec. \ref{sec:numerical}.

With the help of \eqref{euler1}--\eqref{euler3}
derived from \eqref{collision_operator} via an eight
moments ansatz, neglecting hydro motion, and integration over
a finite spherical volume, the following temperature flow model
can be derived:
\begin{align} 
\label{eq:temperaturedynamicsa}
  \frac{3}{2}n_i\frac{\partial kT_i}{\partial t}
  &=\sum_\ell Q_{i\ell}+Q_{\ce{fus \rightarrow i}} \, , \\
\label{eq:temperaturedynamicsb}
  \frac{3}{2}n_e\frac{\partial kT_e}{\partial t}
  &=\sum_\ell Q_{e\ell}+Q_{\ce{fus \rightarrow e}}
  -Q_{\text{rad}}-Q_{e}\, ,
\end{align}
where the index $i$ refers to different ion species, $e$ to
electrons, $n$ to particle densities, and $kT$ to
temperatures. Volume integration leads to boundaries at which
radiation, electronic heat loss, $\alpha$-particle, and neutron
losses can occur. The power densities $Q_{\ce{fus \rightarrow i}}$ and
$Q_{\ce{fus \rightarrow e}}$ are fusion power densities
obtained by considering the fusion energy
deposition due to $\alpha$-particle stopping
as discussed in \ref{sec:alpha-stopping}, due to
$\alpha$-particle loss as discussed in 
\ref{sec:alphaLoss+HeatConduction}, neutron energy deposition
due to elastic scattering as discussed in
\ref{sec:neutron-scattering}, and due to neutron loss at the
boundaries. As we shall show, a significant fraction of the $\alpha$-particle energy
directly heats the ions in non-cryogenic DT at the
required ignition temperatures $kT_{\ce{e}}$. Moreover, the fusion neutrons
pass non-negligible fractions of their energy directly to the
$\ce{DT}$-ions through elastic scattering.

\begin{figure}[ht]
    \centering
    \includegraphics[width=0.35\textwidth]{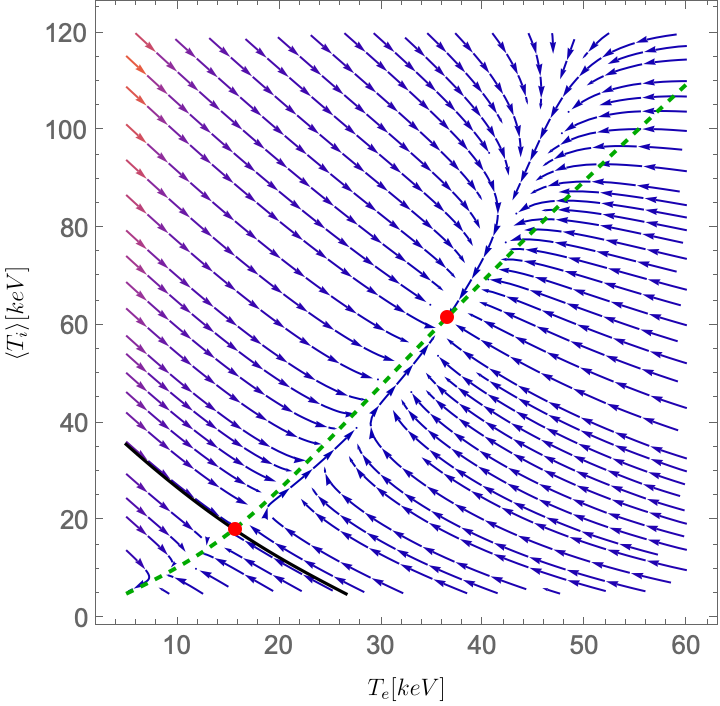} 
    \includegraphics[width=0.35\textwidth]{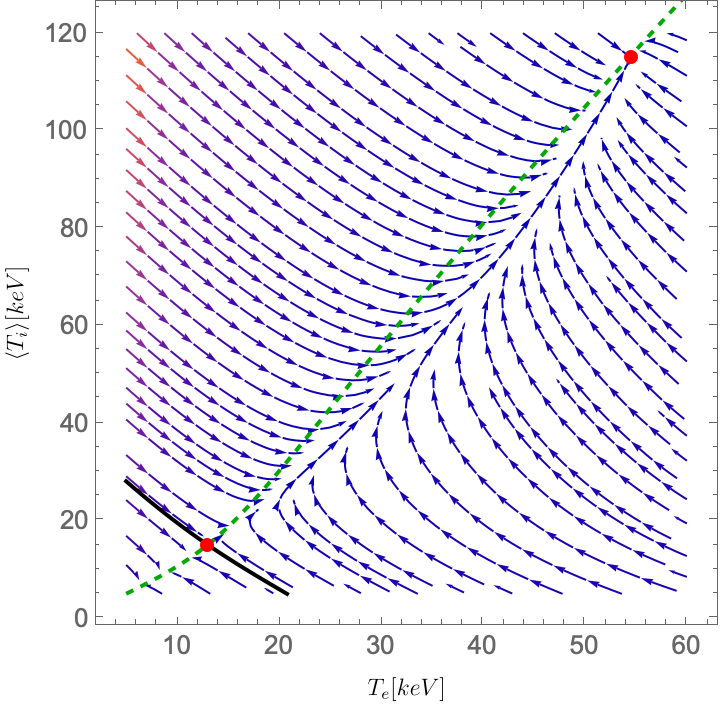}\\
    \caption{Temperature flow diagrams for $\ce{Be[BD_2T_2]_2}$ 
        at $\rho_{\ce{DT}} R=0.5 \, \ce{g} \, \ce{cm^{-2}}$ (top) and $\rho_{\ce{DT}}
        R=1.0 \, \ce{g} \, \ce{cm^{-2}}$ (bottom), where $\rho_{\ce{DT}} = 50 \,
        \ce{g} \, \ce{cm^{-3}}$, $R=0.01 \, \ce{cm}$, and $R=0.02 \, \ce{cm}$.
        The plots illustrate the flow of the electron temperature
        $kT_{\ce{e}}$ and the averaged ion temperature $\langle 
        kT_{\ce{i}} \rangle$ according to Eqs. \eqref{eq:temperaturedynamicsa}
        and \eqref{eq:temperaturedynamicsb}. The important findings are
        (i) $\left \langle kT_{\ce{i}} \right \rangle > kT_{\ce{e}}$
        on the central line and (ii) that ignition happens if the systems
        are started in the regions above the solid black lines in both
        plots. Temperatures higher than those at the upper red dots
        cannot be stably reached according to the model. According to Fig.
        \ref{fig:mixedfuels} and the present figure, higher values of
        $\rho_{\ce{DT}} R$ lower the ignition temperatures and
        increase the peak temperatures the system can reach. For the
        case of $\rho_{\ce{DT}} R = 0.5 \, \ce{gcm^{-2}}$ ignition can
        occur for $kT_{\ce{e}} \approx 16 \, \ce{keV}$ and $kT_{\ce{i}}
        \approx 20 \, \ce{keV}$. For the case of $\rho_{\ce{DT}} R =
        1.0 \, \ce{gcm^{-2}}$ ignition can occur at $kT_{\ce{e}} \approx 14 \,
        \ce{keV}$ and $kT_{\ce{i}} \approx 18 \, \ce{keV}$.}
    \label{fig:MoreauBeBDTrhoR1}
\end{figure}

\paragraph{Ignition diagrams}
Numerically solving \eqref{eq:temperaturedynamicsa} and
\eqref{eq:temperaturedynamicsb} for different initial temperatures
yields the temperature flow diagram
in Fig. \ref{fig:MoreauBeBDTrhoR1}, here shown for heavy beryllium borohydride at
$\rho_\mathrm{DT} R = 0.5 \, \mathrm{g}\,\mathrm{cm}^{-2}$ (top) and
$\rho_\mathrm{DT} R = 1.0 \, \mathrm{g}\,\mathrm{cm}^{-2}$ (bottom).
For illustrative purposes, we represent the flow in the plane of
the electron temperature $kT_e$ and the average ion temperature
$\langle kT_{\ce{i}} \rangle = \sum_i n_i kT_{\ce{i}} / \sum_i n_{\ce{i}}$.
The temperature flows are obtained by plotting the flow directions
obtained from two consecutive numerical temperature values.

The diagram shows that the temperature flow is first attracted to 
the green central line by fast inter-species power equilibration. Thereafter, the temperature flow follows this central line to one of the fixed points.
The central line thus defines a quasi-equilibrium of ion
and electron temperatures, i.e, more precisely, a function $kT_{\ce{i}}(kT_{\ce{e}})$ for every ion species $i$. This quasi-equilibrium can be approximately
determined by considering the flow of the ionic temperatures for a
fixed electron temperature (see the green line in the diagram). Details are given in \ref{sec:equilibrium}. As we can see, the quasi-equilibrium is reached for
higher ion than electron temperatures, especially at overall higher
temperatures. Moreover, as illustrated in Figure
\ref{fig:MoreauBeBDTrhoR1} two fixed points lie on the
quasi-equilibrium line: I) An unstable one at $kT_e \approx 13 \,
\mathrm{keV}$ and $\langle kT_i \rangle \approx 15 \, \mathrm{keV}$
for $\rho_{\ce{DT}} R = 1.0 \, \ce{g} \, \ce{cm^{-2}}$ and II) a stable one at $kT_e
\approx 54 \, \mathrm{keV}$ and $\langle kT_i \rangle \approx 115 \,
\mathrm{keV}$ for the same $\rho_{\ce{DT}} R$. For $\rho_{\ce{DT}} R =
0.5 \, \ce{g} \, \ce{cm^{-2}}$, they move to $kT_e \approx 16 \,
\mathrm{keV}$, $\langle kT_i \rangle \approx  18\, \mathrm{keV}$ and
$kT_e \approx 38 \, \mathrm{keV}$, $\langle kT_i \rangle \approx 65\,
\mathrm{keV}$, respectively.

The stable fixed point
is the point of self-sustained fusion burn. It exists because fuel
consumption is not considered here. The unstable fixed point is the
point in quasi-equilibrium through which the solid black line passes
that separates the region of cooling down to zero temperatures from
the attraction basin of the stable fixed point. It can also be called
the ignition point. However, it is the joint existence of the two fixed
points that shows that ignition is possible. When lowering the
parameter $\rho_{\ce{DT}} R$ (see Fig.
\ref{fig:MoreauBeBDTrhoR1} bottom), they move closer together, merge into a
single point at which the minimal $\left( \rho_{\ce{DT}} R
\right)_{\ce{min}}$ and minimal temperatures $\left( kT_{\ce{e}}
\right)_{\ce{min}}$ and $\left( kT_{\ce{i}} \right)_{\ce{min}}$
are reached and finally disappear.

\begin{figure}[ht]
  \includegraphics[width=0.45\textwidth]{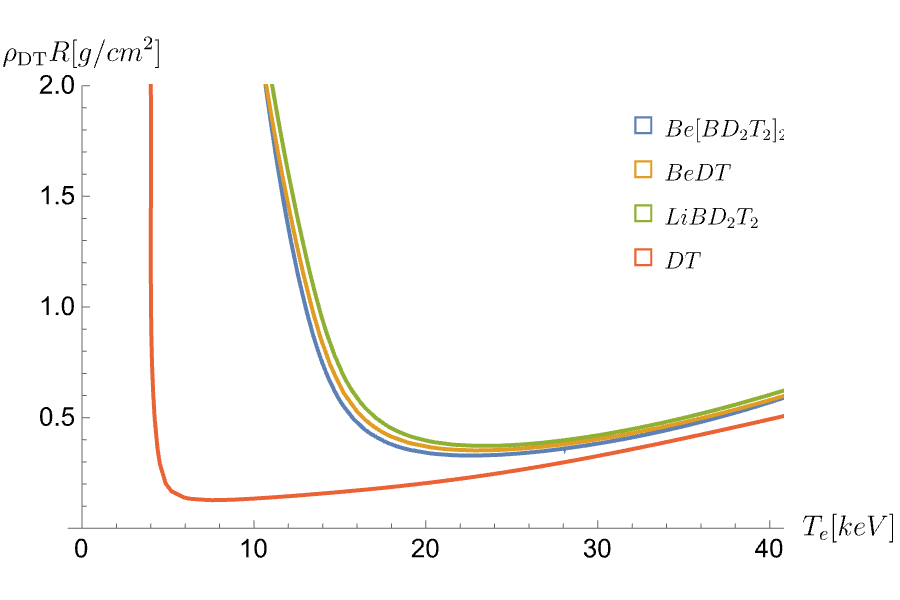}\\
\caption{Ignition conditions for 
$\ce{Be[BD_2T_2]_2}$, $\ce{BeDT}$, $\ce{LiBD_2T_2}$, and $\ce{DT}$.
calculated via the analytical model \eqref{eq:temperaturedynamicsa}
and \eqref{eq:temperaturedynamicsb}. Individual temperatures $kT_{\ce{b}}$ for
each of the constituents $\ce{b}$ in the fuel have been considered.}
\label{fig:mixedfuels}
\end{figure}

\begin{table}[h]
  \small 
    \centering 
    \begin{tabular}{c|c|c}
      compound & \( \left( \rho_{\ce{DT}} R \right)_{\ce{min}}\)
      & \( \left( kT_{\ce{e}} \right)_{\ce{min}}\)\\ 
      \hline
      DT & 0.13 & 7\\
      $\ce{Be}[\ce{BD}_2\ce{T}_2]_2$ & 0.33 & 22\\
      BeDT & 0.35 & 23\\
      $\ce{LiBD}_2\ce{T}_2$ & 0.37 & 23\\
      \hline 
      \end{tabular} 
\caption{Minimal confinement parameters $\left( \rho_{\ce{DT}} R \right)_{\ce{min}}$
and corresponding electron temperatures $kT_{\ce{e}}$ required for
ignition of non-cryogenic DT compounds as obtained from the analytical
model}
\label{tab:ignitionparameters}
\end{table}

The minimal ignition parameters can be studied by setting gain and loss terms (see \ref{sec:gainloss}) equal to each other and plotting the resulting ignition curves in the $kT_{\ce{e}}$-$\rho_{\ce{DT}} R$ plane. Here, it is assumed that the ionic species assume the temperatures $kT_{\ce{i}}(kT_{\ce{e}})$ in quasi-equilibrium. The ignition curves separate the regions of cooling down and of ignition. As visible in Fig. \ref{fig:mixedfuels}, they are flat around their minimum, implying that a small increase in $\rho_{\ce{DT}} R$ leads to a significant decrease in the necessary ignition temperature $kT_{\ce{e}}$.
The minimal confinement parameters $\left(
\rho_{\ce{DT}} R \right)_{\ce{min}}$ and corresponding ignition temperatures $\left( kT_{\ce{e}} \right)_{\ce{min}}$ are listed in Table \ref{tab:ignitionparameters} for the different fuel compounds.
We find that among the non-cryogenic DT fuels considered, heavy beryllium borohydride has the lowest value of $\left( \rho_{\ce{DT}} R\right)_{\ce{min}}$, followed by heavy beryllium hydride, the fuel advocated for by Guskov et al. \cite{gus2011effect}.

\begin{figure*}
\begin{subfigure}{0.63\textwidth}
  \centering
  \includegraphics[width=\textwidth]{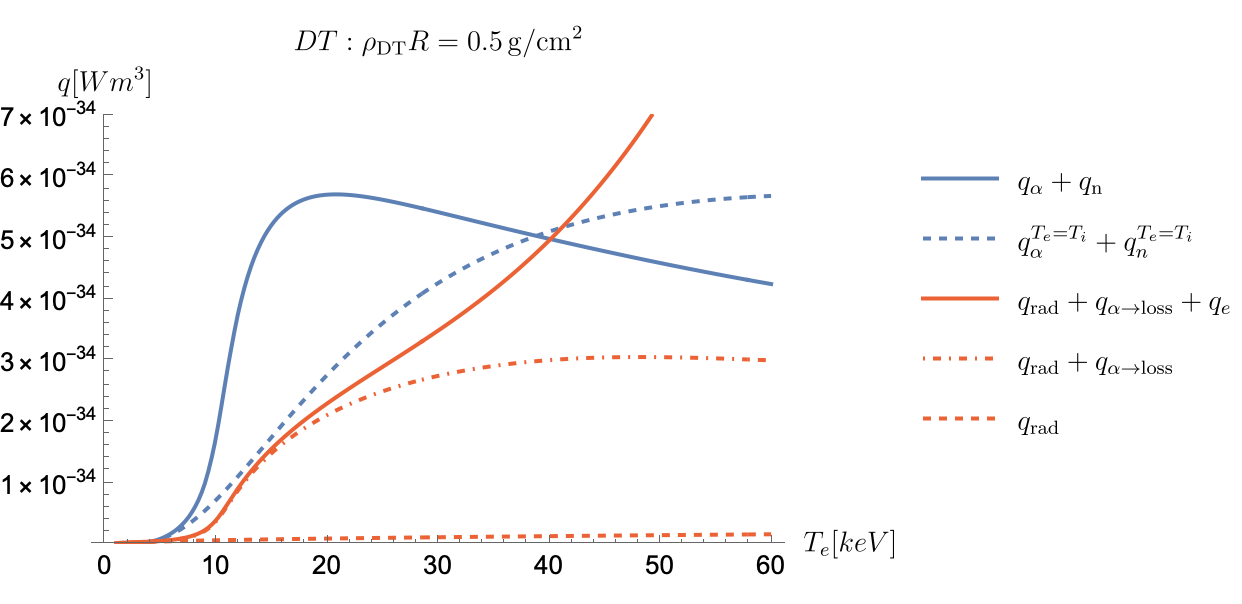}
\end{subfigure}%
\begin{subfigure}{0.27\textwidth}
  \centering
  \includegraphics[width=\textwidth]{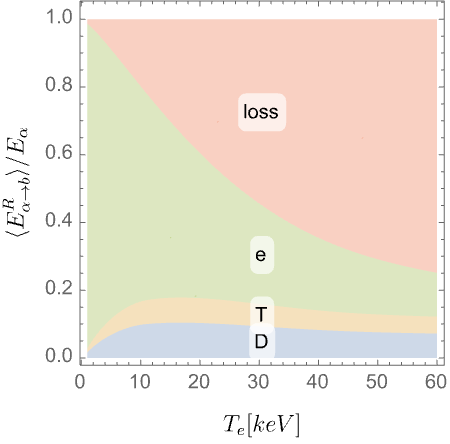}
\end{subfigure}
\\
\begin{subfigure}{0.63\textwidth}
  \centering
  \includegraphics[width=\textwidth]{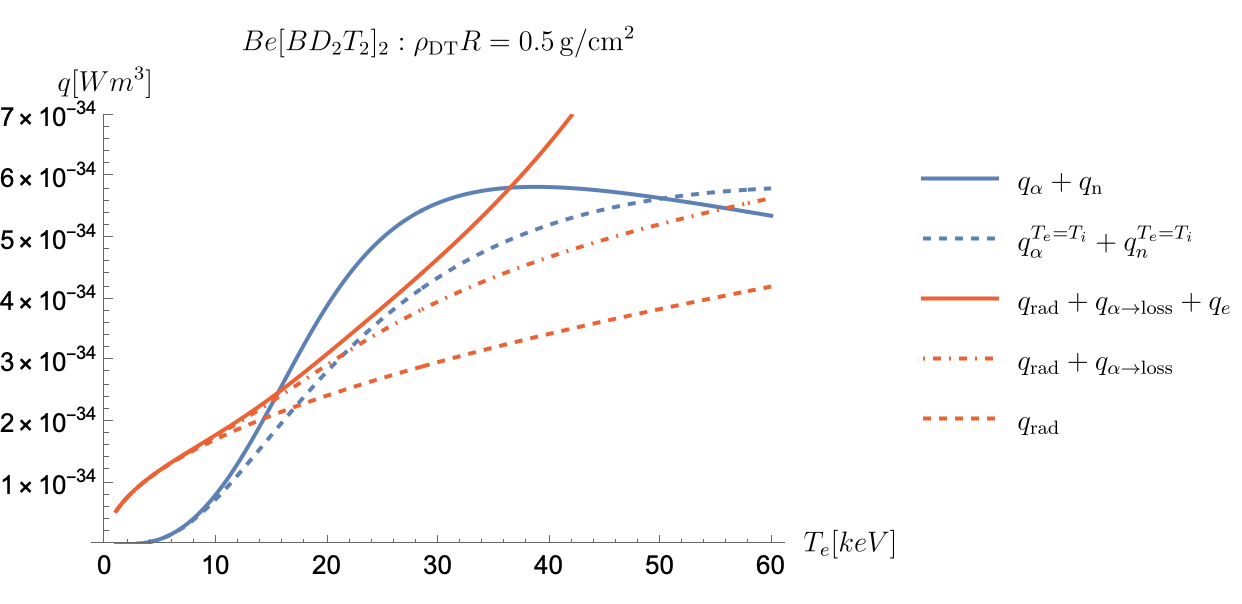}
\end{subfigure}%
\begin{subfigure}{0.27\textwidth}
  \centering
  \includegraphics[width=\textwidth]{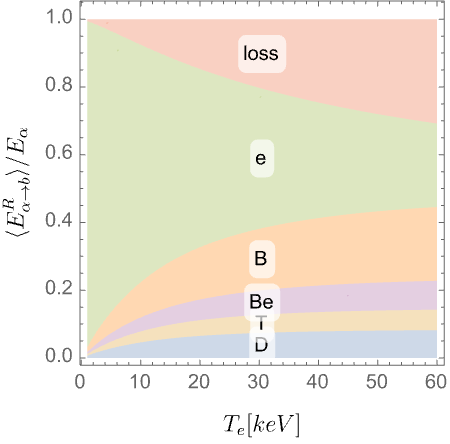}
\end{subfigure}
\\
\begin{subfigure}{0.63\textwidth}
  \centering
  \includegraphics[width=\textwidth]{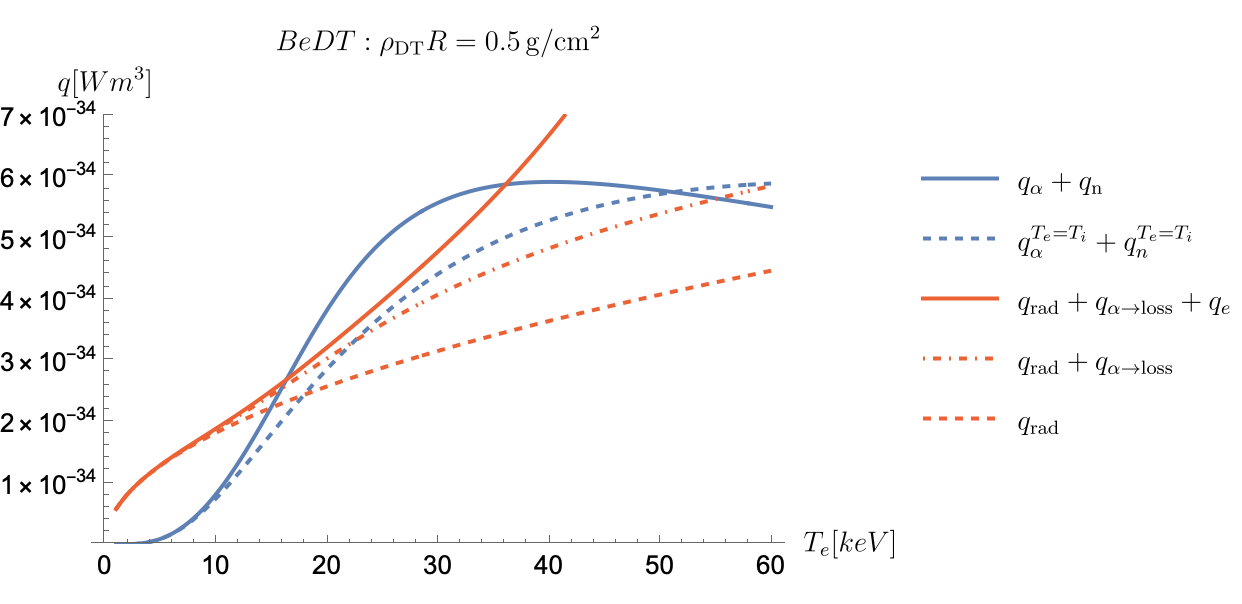}
\end{subfigure}%
\begin{subfigure}{0.27\textwidth}
  \centering
  \includegraphics[width=\textwidth]{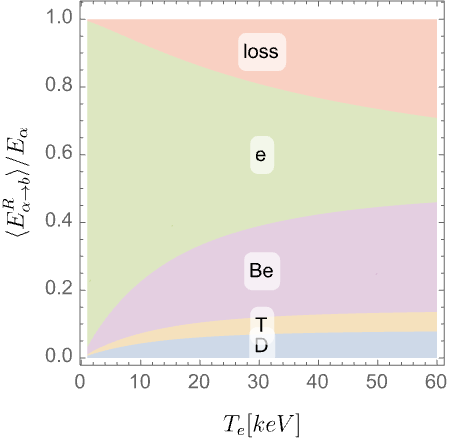}
\end{subfigure}
\\
\begin{subfigure}{0.63\textwidth}
  \centering
  \includegraphics[width=\textwidth]{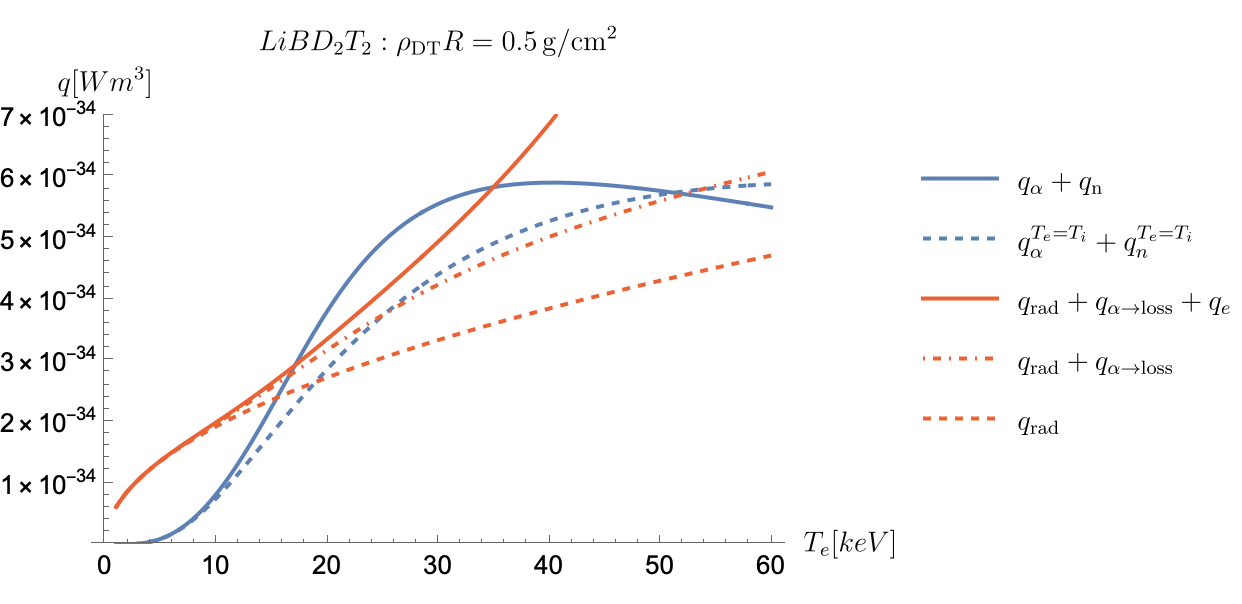}
\end{subfigure}%
\begin{subfigure}{0.27\textwidth}
  \centering
  \includegraphics[width=\textwidth]{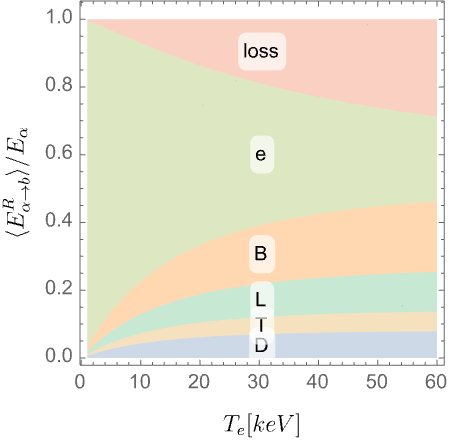}
\end{subfigure}
\end{figure*}

\addtocounter{figure}{-1}
\begin{figure}
  \caption{(Previous page.) Left: Normalized power densities $q=Q/n_{\ce{D}}n_{\ce{T}}$. 
      Comparison of $q_\alpha+q_{\ce{n}}$ for $kT_{\ce{e}}$
      and $\left \langle kT_{\ce{i}} \right \rangle$ in quasi-equilibrium 
      (solid blue line) and $q_\alpha+q_{\ce{n}}$ for $kT_{\ce{e}} =
      kT_{\ce{i}}$ (dashed blue line) with $q_{\ce{rad}}+q_{\ce{\alpha 
      \rightarrow loss}} +q_{\ce{e}}$, $q_{\ce{rad}}+q_{\ce{\alpha
      \rightarrow loss}}$, and $q_{\ce{rad}}$ versus $kT_{\ce{e}}$
      at $\rho_{\ce{DT}} R=0.5 \, 
      \ce{g} \, \ce{cm}^{-2}$, where $q_\alpha$ is the normalized fusion 
      power density in $\alpha$ particles, $q_{\ce{n}}$
      the normalized deposited fusion power density in neutrons,
      $q_{\ce{fus \rightarrow loss}}$ the normalized 
      $\alpha$-particle loss, and $q_{\ce{e}}$ the normalized 
      electronic heat conduction loss. 
      Right: Fraction of $\alpha$-particle 
      energy transferred to the individual constituents of 
      the compound and the electrons versus $kT_{\ce{e}}$ at the 
      corresponding ion temperatures $kT_{\ce{i}}$ in quasi-equilibrium.}
      \label{fig:BeB_channels}
\end{figure}

\paragraph{Role of unequal temperatures for ignition}
Ignition of the fuels in Table \ref{tab:ignitionparameters} is
possible because of two crucial facts: First, $kT_{\ce{i}}
> kT_{\ce{e}}$ facilitates ignition. This is illustrated for
$\ce{Be[BD_2T_2]_2}$ at $\rho_{\ce{DT}} R=0.5 \,\ce{g} \,
\ce{cm}^{-2}$ in Fig. \ref{fig:BeB_channels} (second row left), where
it is shown that for the temperatures $kT_{\ce{i}} > kT_{\ce{e}}$ in
the quasi-equilibrium the system adopts the fusion power is
significantly greater than at equal temperatures $kT_{\ce{e}}=
kT_{\ce{i}}$. In fact, at equal temperatures, the cumulative losses
(solid red curve) always exceed the fusion power (dashed blue curve)
so that ignition would not be possible at this $\rho_{\ce{DT}} R$. The
reason for this behavior is that higher values of $kT_{\ce{i}}$
normally lead to higher fusion power while radiation and electron heat
conduction losses increase with $kT_{\ce{e}}$. Second, the ionic
$\alpha$-particle stopping power of the fuel compound
$\ce{Be[BD_2T_2]_2}$ becomes comparable to the electronic
stopping power at the ignition temperature and eventually
dominates at even higher temperatures (see Fig.
\ref{fig:BeB_channels} second row right). In this way, the
temperature spread between $kT_{\ce{i}}$ and $kT_{\ce{e}}$
is sustained.

We now compare deposition powers to electrons and ions of
non-cryogenic DT fuels ($\ce{Be[BD_2T_2]_2}$, $\ce{BeDT}$ and $\ce{Li
  BD_2T_2}$) with those of pure DT at $\rho_{\ce{DT}} R=0.5 \, \ce{g}
\, \ce{cm}^{-2}$ (see Figure \ref{fig:BeB_channels}). For all these
fuels and for $kT_e < 10 \, \mathrm{keV}$, most of the
$\alpha$-particle energy is deposited into electrons and losses due to
$\alpha$-particles escaping the system are small. For
$kT_e > 20 \, \mathrm{keV}$, however, these losses increase sharply for
$\ce{DT}$ while for $\ce{Be[BD_2T_2]_2}$, $\ce{Li BD_2T_2}$, and
$\ce{BeDT}$, they are largely compensated by the $Z>1$ ions. The
reason for this behavior is that the electronic stopping power
decreases as $kT_e^{-3/2}$ while the ionic one is approximately
independent of temperature in this range. A detailed discussion is
given in \ref{sec:alpha-stopping}. Consequently, the share of
fusion power deposited into electrons decreases with temperature while
the share deposited into ions increases. As mentioned above, this is
a major reason why the ions can sustain significantly greater temperatures than the
electrons. Among the ionic species, most of the fusion power is
deposited to the one with the greatest charge number $Z$, since stopping
power increases with $Z^2$.

Overall, we find that the selected non-cryogenic DT compounds have
significantly greater alpha stopping power than DT. This implies that
the higher radiation loss of non-cryogenic DT compounds is to a great
extent compensated by their enhanced stopping power for
$\alpha$-particles (and neutrons), especially for $kT_{\ce{e}} > 10 \,
\ce{keV}$. The temperature spread between $kT_{\ce{i}}$ and
$kT_{\ce{e}}$ is present for both DT and non-cryogenic DT compounds
and becomes significant for $kT_{\ce{e}} > 10 \, \ce{keV}$. As this is
the temperature range relevant for the ignition of non-cryogenic DT
compounds, it has a stronger impact on the possibility of ignition for
the latter.

\section{Numerical simulations} \label{sec:numerical}
We now compare our analytical model with the predictions of the freely
available numerical hydro code MULTI-IFE, which is well-tested in its
original version for pure DT. For non-cryogenic DTs the code requires
adaptations which we discuss in \ref{sec:multi1D_modified}. By
simulating cases similar to the ones discussed on the basis of the
analytical model, we obtain an approximate cross-check of the results.

In order to create a direct comparison with the temperature
flow analysis of \eqref{eq:temperaturedynamicsa} and
\eqref{eq:temperaturedynamicsb}, the hydro motion in the code
is switched off corresponding to an isobaric scenario. Unlike in the analytical model, fuel consumption and radiation re-absorption are taken
into account in the simulations.

\begin{figure*}[ht]
\centering 
\begin{subfigure}{0.48\textwidth}
\includegraphics[width=\textwidth]{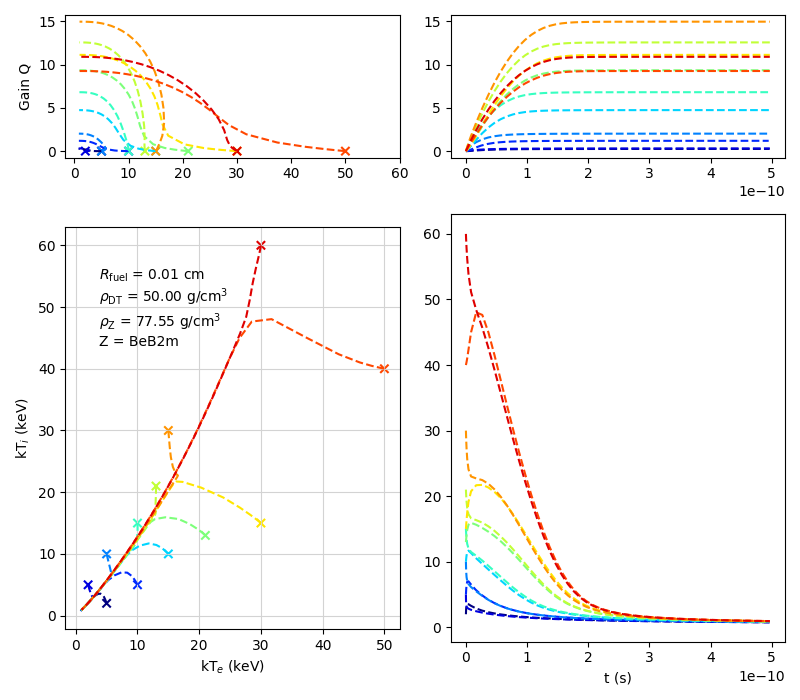}
\end{subfigure}
\begin{subfigure}{0.48\textwidth}
\includegraphics[width=\textwidth]{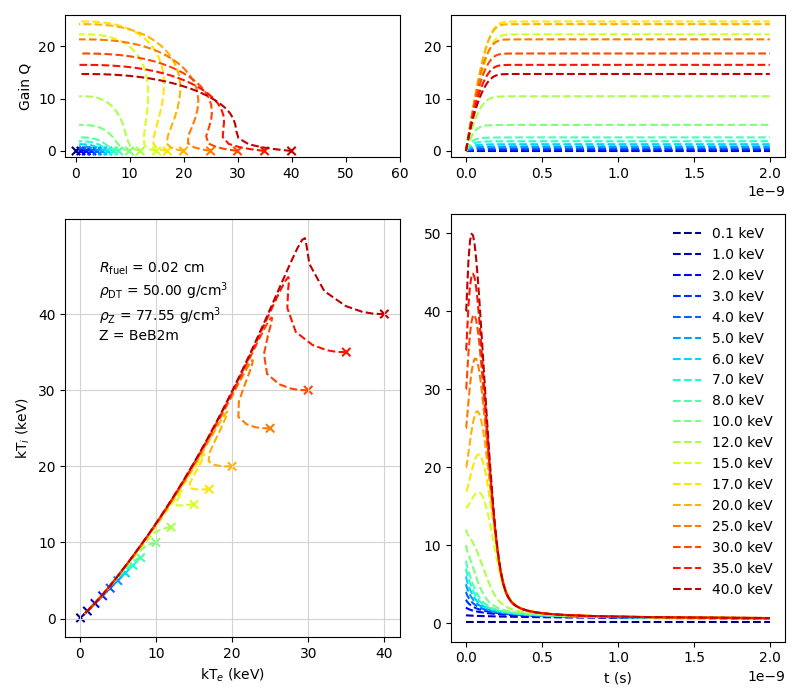}
\end{subfigure}
\caption{Results of two sets of MULTI-IFE simulations  without hydro
  motion. The left four panels correspond to the case where
  $R_\mathrm{fuel}=0.01$~cm and the right four panels for
  $R_\mathrm{fuel}=0.02$~cm. The material in both simulations is
  $\ce{Be[BD_2T_2]_2}$ with the density $\rho_{\ce{DT}} =2\langle\rho_{\ce{DT}}\rangle= 50 \,
   \ce{g} \, \ce{cm^{-3}}$ in the $\ce{DT}$ cells and different initial electron and ion 
    temperatures. These settings correspond to $\langle\rho_{\ce{DT}}\rangle
    R = 0.25 \, \ce{g} \, \ce{cm^{-2}}$ for the left case and
    $\langle\rho_{\ce{DT}}\rangle R = 0.5 \, \ce{g} \, \ce{cm^{-2}}$ for the right
    case since $\ce{DT}$ and $\ce{Z}$ cells alternate. For both cases
    top left: Gain versus electron temperature. Bottom left:
    Trajectories in the $\left( kT_{\ce{e}}, kT_{\ce{i}} \right)$ plane. Top right:
    Gain as a function of time. Bottom right: Ion temperature as a
    function of time. The initial internal energy for the numerical
    setup corresponding to the orange dashed line in the left plot
    with $kT_{\ce{e}} \approx 15 \, \ce{keV}$ and $kT_{\ce{i}} \approx
    30 \, \ce{keV}$ is approximately $494 \, \ce{kJ}$ and for the
    orange dashed line in right plot with $kT_{\ce{e}} = kT_{\ce{i}}
    \approx 20 \, \ce{keV}$ it is about $4.02 \, \ce{MJ}$.
    No or only marginal ignition takes place for the left scenario
    while there is robust ignition for the right one in agreement with
    Fig. \ref{fig:mixedfuels}.}
    \label{fig:report_tite_a1}
\end{figure*}

Several parametric simulations for a highly compressed spherical
$\ce{Be[BD_2T_2]_2}$ plasma at $\langle\rho_{\ce{DT}}\rangle R = 0.25 \,
\ce{g} \, \ce{cm}^{-2}$ with different initial electron and ion temperatures
have been performed, yielding the flow trajectories in the $\left(
kT_{\ce{e}}, kT_{\ce{i}} \right)$-plane shown in Figure
\ref{fig:report_tite_a1} (bottom left). The further plots show the
fusion gain as a function of $kT_{\ce{e}}$ (top left), the gain as a
function of time (top right) and $kT_{\ce{i}}$ as a function of time
(bottom right).

The numerical results agree qualitatively with the predictions of our
analytical model. The simulated temperature flow diagram is similar
to the analytical one in Figure \ref{fig:MoreauBeBDTrhoR1}. The 
simulations also show that a quasi-equilibrium is reached prior
to rising temperatures due to fusion gain. The code predicts
$kT_{\ce{i}} > kT_{\ce{e}}$ even slightly more pronounced than in the
analytical model.

\begin{figure*}[ht]
\centering 
\begin{subfigure}{0.48\textwidth}
\centering
\includegraphics[trim={0 0 0 3cm},clip,width=0.7\textwidth]{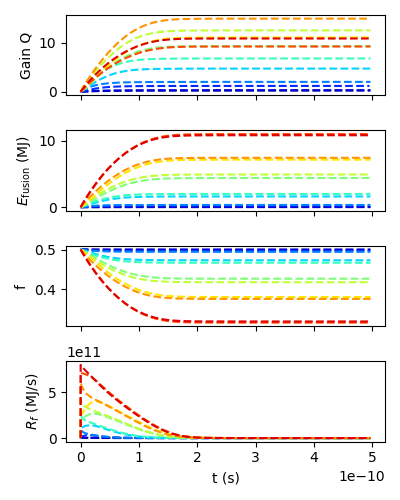}
\end{subfigure}
\begin{subfigure}{0.48\textwidth}
\centering
\includegraphics[trim={0 0 0 3cm},clip,width=0.7\textwidth]{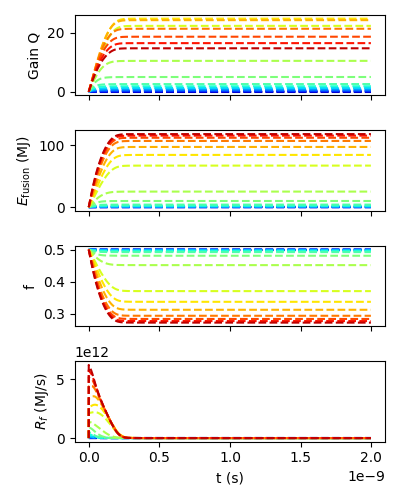}
\end{subfigure}
\caption{
  Results of MULTI-IFE simulations of $\ce{Be[BD_2T_2]_2}$ without hydro
  motion and $\rho_{\ce{DT}} R = 0.25 \, \ce{g} \, \ce{cm}^{-2}$ (left panels)
  and $\rho_{\ce{DT}} R = 0.5 \, \ce{g} \, \ce{cm}^{-2}$ (rightcpanels) for a
  number of initial electron and ion temperatures. The results in this
  figure correspond to the simulations from Figurec\ref{fig:report_tite_a1}.
  Top: Fusion energy versus time. Middle: $f$-number versus time, where
  $f=n_{\ce{D}}(t)/\left( n_{\ce{D}}(t=0) + n_{\ce{T}}(t=0)\right)$. Bottom:
  Fusion power versus time.}
\label{fig:report_tite_a2}
\end{figure*}

Below certain initial temperatures, the system does not ignite but
cools down to zero temperatures while above them further fuel
heating and then cooling sets in. Since the $f$-number (fractional
fuel content) drops (see Fig. \ref{fig:report_tite_a2}), the cooling
is the result of fuel depletion, which contrary to the analytical
model precludes the existence of an upper stable fixed point. The
turning points of the trajectories and the fusion gain depend on the
initial temperatures. The simulations in Fig. \ref{fig:report_tite_a1}
suggest that for $\rho_{\ce{DT}} \, R = 0.5 \, \ce{gcm^{-2}}$
ignition is possible for initial temperatures $kT_{\ce{e}} =
kT_{\ce{i}} = 15 \, \ce{keV}$, which agrees 
well with those predicted by the analytical model for the same
$\rho_{\ce{DT}} \, R$.

Figure \ref{fig:report_tite_a2} complements Fig. \ref{fig:report_tite_a1}
for the same parameters and assumptions. The plots show the fusion
energy as a function of time (top), the fractional fuel content $f(t) = n_{\ce{D}}(t)/(n_{\ce{D}}(0) + n_{\ce{T}}(0))$ as a function of time (middle) and
the fusion power versus time (bottom).

\section{The fusion gain}\label{sec:rescale}
According to Figure \ref{fig:mixedfuels}, ignition requires a
minimum $kT_{\ce{e}}$ at a given $\rho_{\ce{DT}} R$. Assuming equal and homogeneous initial electron and ion
temperatures $kT_{\ce{e}} = kT_{\ce{i}}$ the gain limit of a volume igniter for
$\ce{Be[BD_2T_2]_2}$ can be estimated to be
\begin{eqnarray}
  \label{gain_limit}
  Q 
  &=&
  \frac{2 \, N_{DT}}{3 \left( N_i + N_e \right)} \, \alpha_{DT} \, 
            \frac{\epsilon_{DT}}{kT_e} \\
  &=&
  \frac{2 \, \rho_{DT}}{12
    \, \rho_{DT}  + \frac{255}{31} \, \rho_Z} \, \alpha_{DT} \, 
            \frac{\epsilon_{DT}}{kT_e}
  \nonumber \\  
  &\le&
  \frac{2 \, \rho_{DT}}{12
    \, \rho_{DT}  + \frac{255}{31} \, \rho_Z} \, 
            \frac{\epsilon_{DT}}{kT_e}
   \, , \nonumber 
\end{eqnarray}
where $\alpha_{\ce{DT}}$ is the burn fraction of $\ce{DT}$. It is a
function of $\rho_{\ce{DT}} R$, $\rho_{\ce{Z}} R$, the confinement
time $\Delta t$, the electron temperature $kT_{\ce{e}}$, and the ion
$kT_{\ce{i}}$. Since we are only interested in the gain limit, it
suffices to know that $\alpha_{\ce{DT}} \le 1$ holds. As
Eq. \eqref{gain_limit} shows, the gain limit is lowered by
the inactive constituents of the non-cryogenic DTs. For the case of
$\ce{Be[BD_2T_2]_2}$, implying the natural densities $\rho_{\ce{DT}} \approx 0.31 \,
\ce{g} \, \ce{cm^{-3}}$, $\rho_{\ce{Z}} \approx 0.48 \, \ce{g} \, \ce{cm^{-3}}$ and
$\epsilon_{\ce{DT}} \approx 17.6 \, \ce{MeV}$, the temperature
range sufficient for ignition lies between $12 \, \ce{keV} \le kT_{\ce{e}} \le 22 
\, \ce{keV}$ according to Fig. \ref{fig:mixedfuels} implying $Q \le
120$ according to (\ref{gain_limit}). Larger $\rho_{\ce{DT}} R$
parameters at ignition will bring $Q$ closer to this limit. To obtain
an estimate for $\alpha_{\ce{DT}}$ we use the standard
approximation \cite[eqs. 2.27]{atzeni2004physics}
\begin{align}
\label{burnfraction}
\alpha_{\ce{DT}}&=\frac{\rho_{\ce{DT}} R}{H_{\ce{DTZ}}+\rho_{\ce{DT}} R} \, .
\end{align}
With the help of the simulation shown on the right side of
Fig. \ref{fig:report_tite_a2}, which is closest to the ignition boundary
in Fig. \ref{fig:mixedfuels} we find $f \approx 0.35$ and obtain
$H_{\ce{DTZ}} \approx 1.2 \, \ce{gcm^{-2}}$. This value is
considerably lower than $H_{\ce{DT}}=7 \, \ce{gcm^{-2}}$ used as
the reference value for $\ce{DT}$ in \cite[eqs. 2.29]{atzeni2004physics},
because we consider an isobaric scenario here, while the value of
$H_{\ce{DT}}=7 \, \ce{gcm^{-2}}$ corresponds to an isochoric scenario.

\begin{figure}
    \centering
\includegraphics[width=0.5\textwidth]{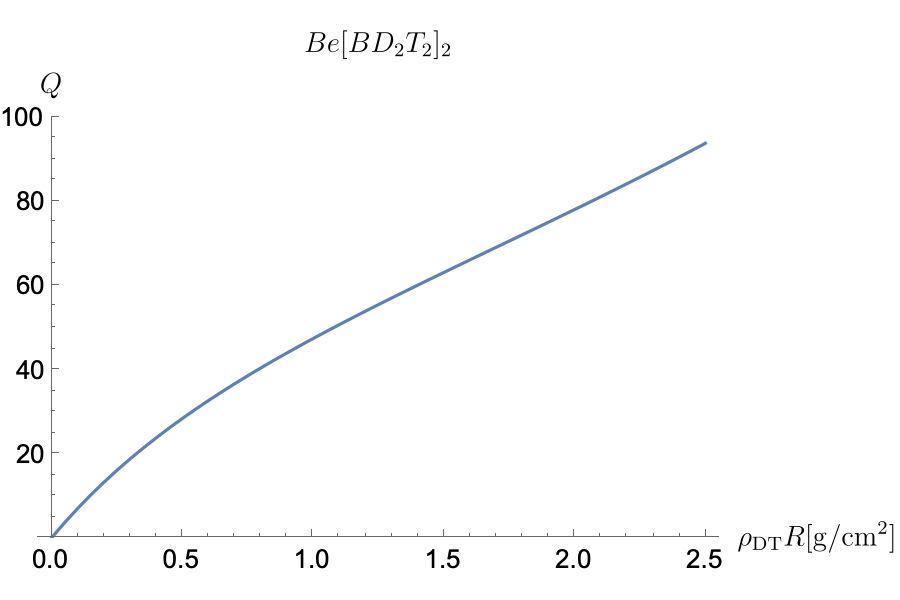}
    \caption{The gain $Q$ as function of $\rho_{\ce{DT}} R$ using \eqref{gain_limit}, \eqref{burnfraction} and linear interpolation for the ignition temperature $kT_e$ obtained from Fig. \ref{fig:mixedfuels}}
    \label{fig:gain}
\end{figure}

Next, we consider the case of the fuel $\ce{Be[BD_2T_2]_2}$ with
$R=0.02 \, \ce{cm}$, $\langle\rho_{\ce{DT}}\rangle R = 0.5 \, \ce{g} \,
\ce{cm^{-2}}$, and $kT_{\ce{e}} = kT_{\ce{i}} = 15 \, \ce{keV}$
given by the right plots in Fig. \ref{fig:report_tite_a1}, which
corresponds to a moderate compression by a factor of $\kappa
\approx 81$. With these parameters the fuel ignites according to
Figs. \ref{fig:mixedfuels} and \ref{fig:report_tite_a1}. The total
initial energy for this case is $E\approx 2.98 \, \ce{MJ}$. 
The total fuel mass consisting of $\ce{DT}$ and $\ce{Z}$ is $M \approx
2.1 \, \ce{mg}$ suggesting the upper limit $E_{\ce{DT}} \approx 282 \,
\ce{MJ}$ for the fusion energy assuming $\epsilon_{DT} \approx 17.6 \,
\ce{MeV}$. The fusion gain predicted by the simulations is $Q \approx
25$, which is in good agreement with Fig. \ref{fig:gain}. According to
Fig. \ref{fig:gain} the fusion gain approaches $Q \approx 100$ for
$\langle\rho_{\ce{DT}}\rangle R \approx 2.5 \, \ce{gcm^{-2}}$. 

It is easy to see that the initial ignition temperature depends on the
product $E \, \rho^2$ for a given $\left \langle \rho_{\ce{DT}} \right
\rangle  R$. We derive a simple expression for the case of
$\ce{Be[BD_2T_2]_2}$. There are
\begin{eqnarray}
  N
  &=&
  \frac{\frac{4\pi}{3} R^3 \rho}{51 m_p}
\end{eqnarray}
fuel molecules, each of which provides $11$ ions and $22$ electrons.
This implies under the assumption that $kT_{\ce{e}} = kT_{\ce{i}}$
holds at ignition
\begin{eqnarray}
  \label{kTe}
  kT_e 
  &\approx&
  \frac{102 \, m_p}{132 \, \pi}
  \frac{E \, \rho^2}{ \left( \rho R \right)^3 } \, ,
\end{eqnarray}
where $\rho = \kappa \, \rho_{\ce{0}}$. For $\rho_{\ce{0}} \approx
0.79 \, \ce{gcm^{-3}}$, $\kappa \approx 81$, $R\approx 0.02 \,
\ce{cm}$ corresponding to $\left \langle \rho_{\ce{DT}} \right \rangle
R \approx 0.5 \, \ce{gcm^{-2}}$, and $E \approx 2.98 \, \ce{MJ}$ the
ignition temperature is $kT_{\ce{e}} \approx 15 \, \ce{keV}$ in
agreement with the iginition temperature in right simulation in
Fig. \ref{fig:report_tite_a1}. As (\ref{kTe}) shows increasing the
fuel density implies lower ignition energy. For $\kappa \approx 324$
we obtain $E \approx 186 \, \ce{kJ}$.

\section{Summary} \label{sec:summary}
  In the present paper it has been explored under which conditions   
  non-cryogenic DTs are capable of igniting and how their fusion
  gain scales for volume ignition. As is shown they can yield
  sufficient fusion gain for fusion energy production. One advantage of
  non-cryogenic DTs is that they do not require cryo-technology.
  However, this feature is only one interesting aspect about
  them. They may also enable potentially simple new fusion reactor
  designs.
  
  Compared to previous work on non-cryogenic DT fuels
  \cite{gus2011effect,gus2013compression,gus2015fast,gus2016influence},
  we have identified two new important positive effects that improve the
  predictions for ignition of these fuels and their usage for energy
  production: 1. Higher ionic than electronic temperatures sustained
  by 2. significant ionic stopping power in the same order of
  magnitude as or even exceeding the electronic one. Moreover, with
  heavy beryllium borohydride we have identified a non-cryogenic DT
  fuel that ignites more easily than the previously best candidate,
  heavy beryllium hydride (BeDT).

  Volume ignition represents a worst case scenario. Non cryogenic DTs
  ignite robustly. According to Fig. \ref{fig:gain} the fusion gain
  approaches $Q \approx 100$ for larger $\langle\rho_{\ce{DT}}\rangle
  R$. The investigation of specific $Q > 100$ scenarios with
  non-cryogenic DTs is left for future work. 

\section{Acknowledgements} \label{sec:ack}
The present work has been funded by Marvel Fusion. Specifically,
motivated by the analysis in the present paper Rafael
Ramis Abril has adapted the stopping power model for
$\alpha$-particles in the original MULTI-IFE code. In addition,
neutronic stopping has been added to the code. However, the
updated MULTI-IFE code has not been used in the present paper.
The consideration of a wider class of fuels beyond DT for fusion
energy production has been suggested by Hartmut Ruhl and
Georg Korn. The technical details have been worked out by
Christian Bild, Matthias Lienert, Markus N\"oth, and Hartmut Ruhl.
Ondrej Pego Jaura has carried out the simulations. Rafael Ramis
Abril has been supported by projects: PID2022-137339OB-C22 of the
``Plan Estatal 2021-2023" of the Spanish Goverment and by
ENR-IFE.01.CEA of EUROFUSION.

\bibliographystyle{elsarticle-num} 
\bibliography{literatur_eqn_motion}

\appendix

\section{Analytical Model} \label{sec:analyticalmodel}
Here we describe the details of our analytical model
\eqref{eq:temperaturedynamicsa}, \eqref{eq:temperaturedynamicsb}. In
\ref{sec:gainloss}, we give the formulas for the loss and gain terms
included in the model. Mainly, the deposition power term needs further
work. In \ref{sec:alpha-stopping}, we introduce our stopping power
model, based on a calculation of friction and heat exchange terms in
the Euler equations. In \ref{sec:alphaLoss+HeatConduction}, we
consider losses of alpha particles through the
boundaries. \ref{sec:neutron-scattering} deals with neutron
scattering. In \ref{sec:equilibrium}, we determine the
quasi-equilibrium of temperatures that has been used for the plots in
sec. \ref{sec:ignition}. In \ref{SEC:disofdis}, we calculate the
distribution of flight distances inside a sphere for random initial
positions and flight directions. \ref{sec:mixedtempreactivity} reduces
the reactivity for two ion species with different temperatures to the
reactivity for a single effective temperature.

\subsection{Gains and losses}\label{sec:gainloss}

Here we specify the different gain and loss mechanisms that are taken
into account in our analytical model \eqref{eq:temperaturedynamicsa},
\eqref{eq:temperaturedynamicsb}.

The most important loss channel is radiation due to bremsstrahlung,
given by \cite[Eq. 4.24.4]{wesson2004tokamaks}
\begin{eqnarray}
  \label{loss:rad}
  Q_{\mathrm{rad}}
  &=&-\frac{2^{1/2} e^6}{3 \,\pi^{5/2}\epsilon_0^3 c^3 h\, m_e^{3/2}}
  \, n_e \sum_i Z_i^2 n_i \sqrt{kT_e}
\end{eqnarray}
with $h$, $e$, $\epsilon_{\ce{0}}$, $c$, and  $m_{\ce{e}}$ being
Planck's constant, the elementary charge, the vacuum permittivity, the
speed of light and the electron mass, respectively.

Next, we consider heat conduction losses. We take them into account
using the formula \cite[Eq. 4.10]{atzeni2004physics}
\begin{align}
  \label{loss:conduction}
    Q_{\ce{e}}&=-\frac{S \chi_{\ce{e}} \nabla kT_{\ce{e}}}{V}\approx-0.5\frac{3\chi_{\ce{e}}
         kT_{\ce{e}}}{R^2} \, , \\
    \chi_e&=\frac{ 2^8
            \epsilon_0^2(kT_e)^{5/2}\sqrt{2\pi}}{\sqrt{m_e}q_e^4\sum_j
            n_j}\sum_i\frac{n_i}{(3.3+Z_i)\ln \Lambda_{ei}} \, ,
\end{align}
where we have chosen the value $0.5$ for the constant $c_{\ce{e}}$ of
\cite{atzeni2004physics} in accordance with \cite[section
2.4]{karlchenvomIPP6}. For the Coulomb logarithm $\ln \Lambda$ we use
\cite[p. 34]{richardson20192019} \footnote{For the interactions of alpha particles
with ions, formula (d) for counter-streaming ions in \cite[p. 34]{richardson20192019} is used. For the electron ion interaction, formula (b) in \cite[p. 34]{richardson20192019}, it can happen that none of the conditions is satisfied. In such a situation, we use the second case of (b).}. Our numerical analysis also takes the free streaming limit into account. It is given by
\begin{align}
    Q_{\ce{e}}= - 0.1 \, \frac{S }{V} n_e kT_e \sqrt{\frac{kT_e}{m_e}}
  \, .
\end{align}
However, this limit only becomes relevant at very high temperatures. 
Both bremsstrahlung and heat conduction losses affect only the electron temperature.

The heat transfer according to \cite{spitzer2006physics} is given by
\begin{align}
\label{eq:Q_bc}
    Q_{bc}=\frac{  q_b^2 q_c^2 n_b n_c \ln
  \Lambda_{bc}}{(2\pi)^{3/2}\epsilon_0^2 m_b m_c }
  \frac{kT_c-kT_b}{\left(\frac{kT_b}{m_b}+\frac{kT_c}{m_c}\right)^{3/2}}
  \, .
\end{align}

Concerning the gain term, we write the overall rate of energy transferred from fusion events to species $b$ as
\begin{align}
  \label{fusionpower}
  Q_{\mathrm{fus}\rightarrow b}
  &= n_D n_T \langle \sigma v_{DT} \rangle 
  \left(\langle E^R_{\alpha\rightarrow b}\rangle 
  +\langle E^R_{n\rightarrow b}\rangle\right)\, ,
\end{align}
where $\langle E^R_{\alpha\rightarrow b}\rangle$ and $\langle E^R_{n\rightarrow b}\rangle$ is the expected
energy deposited into species $b$ by an alpha particle and neutron, respectively. The former will be determined in \ref{sec:alphaLoss+HeatConduction} taking into account stopping power and alpha particles leaving the system due to geometric constraints (see Eq. \eqref{loss:alpha}). The expected deposited energy by neutrons is calculated analogously. In \ref{sec:neutron-scattering} it is explained how to modify the expression obtained for alpha particles appropriately. For the reactivity $\sigma v_{DT}$, we use the interpolation
formula in \cite{bosch1992improved} with the temperature chosen
according to \eqref{mixtemp}.

\subsection{$\alpha$-stopping}\label{sec:alpha-stopping}
Here we present a method to determine how the energy of the fusion
alpha particles is distributed to the different species of charged
particles. We assume a homogeneous spherical plasma of electrons,
deuterium, tritium, and possibly other ions.

Each alpha particle is created with kinetic energy
\(E_\alpha=3.5 \, \mathrm{MeV}\) in the center of mass frame of a pair of
deuterium and tritium ions with a uniformly random direction. The
velocities of the center of mass frames are distributed thermally with
a temperature between the temperatures of the two hydrogen
species. The alpha particle energy \(3.5 \, \mathrm{MeV}\) is
overwhelmingly larger than the thermal energies of deuterium and
tritium. Hence, the resulting distribution in velocity space at the
time of alpha creation is a thin spherical shell with radius \(\sqrt{2
  E_\alpha/m_\alpha}\) and shell thickness corresponding to the spread
in center of mass frames of pairs of deuterium and tritium ions. The
precise hydrodynamics of the system containing alpha particles
generated according to this distribution, their interaction with the
electrons and other ions in the plasma and their influence on the
subsequent generation of alpha particles is complicated and only
tractable via simulations. In order to generate a semi-analytic model
which can be evaluated quickly, we exploit that the density of alpha
particles in the plasma is very low until a significant part of the
plasma has reacted. Moreover, the density of alpha particles in the
\(\mathrm{MeV}\)-range is low at all times. We follow the alpha
particles which are generated at an arbitrary instant. This allows us to
dissect the initial velocity space distribution of the alpha particles
into many small spherical sectors, each with a mean velocity close to
the radius of the sphere and velocity spread close to a thermal
distribution with temperature close to the one of the hydrogen
species. We take the action of the non-alpha species on the alpha
particles into account but neglect any change of properties of the
non-alpha species due to the small densities of alpha particles. We
also neglect interactions between groups of alpha particles for the
same reason.

To derive a one dimensional gain and loss model, we start from the
Fokker-Planck equations. Nuclear collision with ions are neglected, even so they increase the ionic stopping power \cite{ghosh2007energy}. Its collision operator are
given by
\begin{eqnarray}
\label{collision_operator}
C(f_b,f_c)&=&\frac{ q_b^2q_c^2 \ln \Lambda}{8\pi\epsilon_0^2}
\int d^3p_c \, \left(\frac{\partial}{\partial p^k_b}
-\frac{\partial}{\partial p^k_c}\right) \\
&&\times \, \frac{|\bold{v}|^2_\text{rel}
\delta^k_m-v^k_\text{rel}v^m_\text{rel}}{|\bold{v}_\text{rel}|^3}\left(\frac{\partial}{\partial 
p^m_b} - \frac{\partial}{\partial p^m_\ell}\right) \, f_b f_c \, . \nonumber
\end{eqnarray}
We use an approximation of the eight-moment model of \cite[Section
26]{Burgers1969}, \cite[Section H]{generalizedBraginskii}, and
\cite[Equation (73)]{atzeni1987physical}. This means we employ the
following ansatz for the distribution function to plug into \eqref{collision_operator}
\begin{eqnarray}
  \label{LTE}
  f_b
  &=&\frac{n_b}{\left(2\pi
    m_bkT_b\right)^{3/2}}\exp\left(-\frac{(\bold{p}_b-m_b\bold{u}_b)^2}
      {2m_bkT_b}\right) \\
  &&\times \, \left[ 1-  \frac{m_b}{n_b (kT_b)^2} \, 
  \bigg(1-\frac{(\bold{p}_b-m_b\bold{u}_b)^2}{5 m_b T_b} \bigg)
     \right. \nonumber \\
  && \hspace{1cm} \left. \times \, \bold q_b \cdot \left( \frac{\bold{p}_b}{m_b}
     - \bold{u}_b \right) \right] \, . \nonumber
\end{eqnarray}
The heat flux \(\bold q_b\) is then taken to be zero for the analytical model, because we assume a homogeneous density and temperature. The numerical study treats inhomogeneous fluids by solving the corresponding dynamical equation for the heat flux which follows from \eqref{LTE} by a quasi static ansatz.
This implies the following Euler equations for multiple species
with general mean velocities
\begin{eqnarray}
\label{euler1}
\frac{d n_b}{dt}
&=& -n_b \, \frac{\partial}{\partial\bold{r}} \cdot \bold{u}_b \, , \\
\label{euler2}
m_b n_b \, \frac{d {\bf u}_b}{dt}
&=&-\frac{\partial P_b}{\partial\bold{r}}+\sum_\ell\bold{R}_{b\ell} \, ,\\
\label{euler3}
\frac{3}{2} \, n_b \, \frac{d kT_b}{dt}
&=&-P_b \, \frac{\partial}{\partial\bold{r}} \cdot \bold{u}_b 
-\frac{\partial}{\partial\bold{r}} \cdot \bold{q}_b
+\sum_\ell Q_{b\ell} \, .
\end{eqnarray}
The vector $\bold{R}_{b\ell}$ describes the friction between two
species, and $Q_{b\ell}$ the heat exchange. The pressure \(P_b=n_n k T_b\) satisfies the ideal gas law. Neglecting the heat flux \(\bold q_b\), the friction $\bold{R}_{b\ell}$ and  heat exchange $Q_{b\ell}$ are given by 
\begin{eqnarray}
\label{frictionforce}
\bold{R}_{bc}&=&\int d^3p_b \, \bold{p}_b \, C(f_b,f_c) \\
&=&-\frac{q_b^2 q_c^2 n_b n_c \ln \Lambda_{bc}}{(2\pi)^{3/2} \epsilon_0^2 \mu_{bc} } \, 
\frac{\bold{u}_{bc} }{u_{bc}^3} \nonumber \\
&&\times \, \Psi\left(\frac{u_{bc}}{
\sqrt{2\left(\frac{kT_b}{m_b}+\frac{kT_c}{m_c}\right)}}\right) \nonumber
\end{eqnarray}
and
\begin{eqnarray}
\label{Q_bc}
Q_{bc}&=&\int  d^3p_b \, \frac1{2 m_b}(\bold{p}_b-m_b\bold{u}_b)^2 \, C(f_b,f_c) \\
&=&
\frac{q_b^2 q_c^2 n_b n_c \ln \Lambda_{bc}}{ m_b 
    (2\pi)^{3/2}\epsilon_0^2 u_{bc}} \nonumber \\
&&\times \, \Phi\left(\frac{ 
   u_{bc}}{\sqrt{2\left(\frac{kT_b}{m_b}+\frac{kT_c}{m_c}
   \right)}},\frac{
   kT_b}{\mu_{bc}\left(\frac{kT_b}{m_b}+\frac{kT_c}{m_c}\right)}\right) 
   \, . \nonumber 
\end{eqnarray}

These quantities  are obtained as the first and second moments of the
Fokker-Planck collision operator (\ref{collision_operator}) assuming
local thermal equilibrium \eqref{LTE}, cf. \cite{Burgers1969}. The relative velocity is denoted by
\begin{equation}
    \bold{u}_{bc} = \bold{u}_{c}-\bold{u}_{c}.
\end{equation}
The functions $\Psi$ and $\Phi$ in (\ref{frictionforce}) and
\eqref{Q_bc} are given by
\begin{eqnarray}
\Psi(x)&=&\sqrt{\frac{\pi}{2}}   \mathrm{erf}\left(x\right) 
-\sqrt{2}x\exp\left(-x^2\right) \, , \\
\Phi(x,y)&=&\sqrt{\frac{\pi}{2}}   \mathrm{erf}\left(x\right) 
-\sqrt{2}y x\exp\left(-x^2\right) \, . 
\end{eqnarray}
The friction force has two interesting limiting cases given by
\begin{equation}\label{approx}
    \frac{1}{u_\alpha^2}\Psi\left(\frac{u_\alpha}{
    \sqrt{\frac{2kT_b}{m_b}}}\right)\approx\left\{\begin{matrix}\frac{1}{3}\left(\frac{kT_b}{m_b}\right)^{-3/2}
    u_\alpha, &u_\alpha\ll\sqrt{\frac{kT_b}{m_b}}\\
    \sqrt{\frac{\pi}{2}} \, u_\alpha^{-2}, &u_\alpha\gg\sqrt{\frac{kT_b}{m_b}}\, .
    \end{matrix}\right.
\end{equation}
For low temperatures, the friction with electrons is dominant because
of the reduced mass in the denominator of \eqref{frictionforce}. Since
for electrons the first case of \eqref{approx} is relevant, their
contribution to stopping is reduced when the temperature
increases. This increased electron transparency for alpha particles at
higher temperatures is a well-known result. For ions on the other
hand, the second case of \eqref{approx} is relevant where the friction
is independent of temperature. It is worth noting that the friction
term \eqref{frictionforce} is proportional to $Z^2$. Hence, at higher
temperatures or if medium or high-$Z$ materials are present, the ion
contribution to alpha stopping starts to play an important role.

Energy conservation is given by 
\begin{align}
    Q_{bc}+Q_{c b}=\bold{R}_{bc}\cdot \bold{u}_{bc}\,. 
\end{align}
Accordingly, $Q_{bc}$ describes two effects. First, it describes how
the heat that is generated by friction is distributed between both
species. Second, if the two temperatures differ, then heat flows
between those species. In the limit of $ u_{b\ell}\rightarrow 0$
the term $Q_{bc}$ reduces to \eqref{eq:Q_bc}. It is the goal of this section to
determine the frictional heat deposition in alpha particles and
the other species and the subsequent temperature flow from the alpha
particles to the other species. To this end, we solve the hydrodynamic
equations in Lagrangian coordinates, neglecting the pressure terms, as
friction has a much greater effect for such high relative velocities:

\begin{align}
    \frac{d u_\alpha}{d t}=\sum_\ell\frac{R_{\alpha \ell}}{m_\alpha n_\alpha} \, , \\
    \frac{d (kT_\alpha)}{d t}=\sum_\ell \frac{2Q_{\alpha \ell}}{3n_\alpha} \, .
\end{align}
Since all non alpha species are on average at rest, $ u_{\alpha
  \ell}=u_\alpha=\sqrt{\frac{2E_\alpha}{m_\alpha}}$. We parameterize
the dynamics with respect to flight distance instead of the time and
we use the energy instead of the velocity, arriving at
\begin{eqnarray}
\label{dEdr}
\frac{d E_{\alpha}}{d x}
&=&\sum_\ell \frac{- m_\alpha q_\alpha^2 q_\ell^2 n_\ell  \ln
    \Lambda_{\alpha \ell}}{ 2(2\pi)^{3/2} \epsilon_0^2 \mu_{\alpha
    \ell} E_\alpha } \\
&&\times \,\Psi\left(\sqrt{\frac{E_\alpha }{
   m_\alpha\left(\frac{kT_\alpha}{m_\alpha}+\frac{kT_\ell}{m_\ell}\right)}}\right)\nonumber\\
\frac{d kT_{\alpha}}{d x}&=&\sum_\ell
\frac{  q_\alpha^2 q_\ell^2 n_\ell \ln \Lambda_{\alpha \ell}}{ 3
                             (2\pi)^{3/2} \epsilon_0^2 E_\alpha } \\
&&\times \,\Phi\left( \sqrt{\frac{ E_\alpha}{m_\alpha\left(\frac{kT_\alpha}{m_\alpha}
+\frac{kT_\ell}{m_\ell} \right)}}, \frac{ kT_\alpha }{\mu_{\alpha
   \ell}\left(\frac{kT_\alpha}{m_\alpha}+\frac{kT_\ell}{m_\ell}\right)}\right)\nonumber
\end{eqnarray}
These are two coupled differential equations, which depends on the
constants $kT_\ell$. We integrate them up to a distance
\(\min(s,s_\alpha)\). Here, \(s\) is the distance the alpha particles
travel until they leave the fuel and \(s_\alpha\) is the distance
until they lose $99\%$ of the sum of their kinetic energy and the
difference of their initial and final thermal energies.

The criterion to determine the exact value of $s_\alpha$ has a weak
influence on the energy distribution. When the alpha particles are
almost at rest, their kinetic energy decreases exponentially. Hence,
the stopping process is at no point in time concluded. During this
process, the alphas continuously act as thermal bridge between the
other species. Thus, some choice when to stop the dynamics has to be
made. We use the $99\%$ criterion which has the additional advantage of avoiding
numerical instabilities at the end of the dynamics. The amount of heat transferred to the non alpha species
$E_b=\frac{3}{2}n_bkT_b$ per alpha particle can be obtained from the
following equations.
\begin{eqnarray}
\frac{d E_{b}}{dx} 
&=&
\frac{ m_\alpha  q_b^2 q_\alpha^2 n_b \ln \Lambda_{b \alpha}}{ m_b 2(2\pi)^{3/2} \epsilon_0^2 E_\alpha} \\
&&\times \,\Phi\left( \sqrt{\frac{
   E_\alpha}{m_\alpha\left(\frac{kT_b}{m_b}+\frac{kT_\alpha}{m_\alpha}
   \right)}}, \frac{
   kT_b}{\mu_{b\alpha}\left(\frac{kT_b}{m_b}+\frac{kT_\alpha}{m_\alpha}\right)}\right),\nonumber\\
\label{Ealpha->b}
E^s_{\alpha\rightarrow b}
&=&
\int_0^{\min(s,s_\alpha)} dx \, \frac{d E_{b}}{dx} \, .
\end{eqnarray}
We solve these equations numerically for a given state of the fuel 
and a distance \(s\) for the analysis in the main text.

\begin{figure}[ht]
    \centering 
    \includegraphics[width=0.45\textwidth]{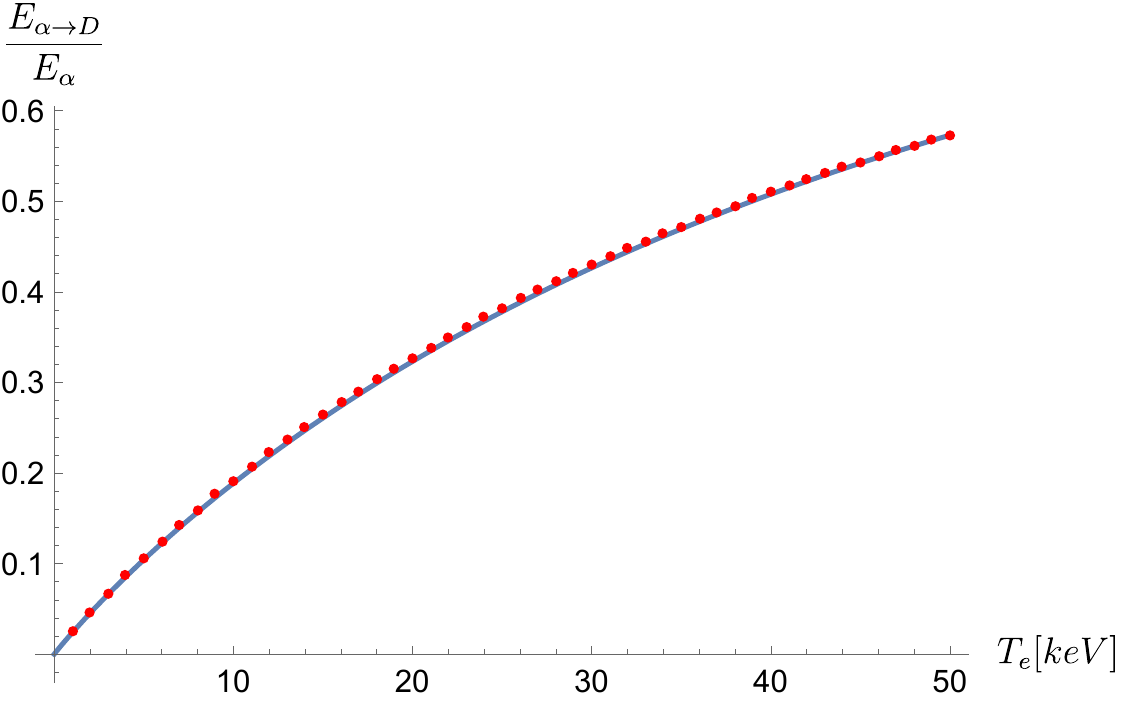}
    \caption{Energy transfer to deuterium ions via $\alpha$-stopping 
    in a deuterium electron plasma for $kT_{\ce{e}} = kT_{\ce{D}}$
    calculated with the model in this paper (red dots) and the model in \
    \cite[Eq. 5.4.12]{wesson2004tokamaks} (blue line).}
    \label{fig:wessoncompare}
\end{figure}

In order to compare our model with the literature, we note that
\cite[chapter 5.4]{wesson2004tokamaks} contains a similar model but
only for the special case of equal ion and electron temperatures. The
energy transfer to ions for the two models is compared in
Fig. \ref{fig:wessoncompare} for a deuterium electron plasma with
equal temperatures, showing excellent agreement.

The next section discusses the losses of due to alpha particles
leaving the plasma.

\subsection{$\alpha$-losses}\label{sec:alphaLoss+HeatConduction}
We now assess the alpha losses associated with finite size,
considering a homogeneous spherical plasma with radius $R$.

Some alpha particles will leave the plasma before they are fully
stopped. To estimate the associated energy loss, we assume that the
generation of alpha particles is equally likely everywhere inside the
sphere and that every flight direction is equally likely. In order to
calculate the energy deposition, it is sufficient to know the
distribution $p(s)$ of the lengths of the trajectories $s$
inside the plasma. It will be determined in \ref{SEC:disofdis}, with
the result \eqref{sprob}. For any length \(s\), the deposited energy is given
by Eq.\eqref{Ealpha->b}. Then the mean deposited energy is given by
the average of that expression, i.e., by
\begin{align}
    \langle E^R_{\alpha\rightarrow b}\rangle=\int_0^{2R} ds \,p(s) 
    \int_0^{\min(s,s_\alpha)} \frac{d E_{b}(E_\alpha(x))}{dx} dx .
\end{align}
Using partial integration, we simplify this expression to
\begin{align}
   \langle E^R_{\alpha\rightarrow b}\rangle &=  \int_{0}^{\min(2R,s_\alpha)}
   \frac{d E_{b}(E_\alpha(x))}{dx}
    dx\nonumber\\
    &-
    \int_{0}^{\min(2R,s_\alpha)} \frac{s(12 R^2 - s^2)}{16 R^3} 
    \frac{d E_{b}(E_\alpha(s))}{dx} ds\\\nonumber
    &= 
     E^\infty_{\alpha\rightarrow b}
    -
    \mathds{1}_{2R < s_\alpha}
    \int_{2R}^{s_\alpha}
   \frac{d E_{b}(E_\alpha(x))}{dx}
    dx\\
    &-
    \int_{0}^{\min(2R,s_\alpha)} \frac{s(12 R^2 - s^2)}{16 R^3} 
    \frac{d E_{b}(E_\alpha(s))}{ds} ds\\\label{loss:alpha}
    &= E^\infty_{\alpha\rightarrow b}-\langle L^R_{\alpha\rightarrow b}\rangle
\end{align}
where $E^\infty_{\alpha\rightarrow b}$ is given by \eqref{Ealpha->b}
corresponding to an infinitely extended plasma with no alpha losses.

The first of the two loss terms corresponds to the energy that an
alpha particle would transfer to the species \(b\) if it could stay
inside the plasma even after traversing a distance of \(2R\). This
term only contributes for \(2R< s_\alpha\). The second term never
vanishes as it includes losses due to outward-moving alpha particles that are
created close to the boundary of the plasma. The overall rate of
energy transferred from alpha particles created in fusion events to species \(b\)
\eqref{fusionpower} can be split into a gain and a loss term:

\begin{align}
    Q_{\mathrm{fus}\rightarrow b}^\alpha= n_D n_T \langle \sigma v_{DT}\rangle 
    ( E^\infty_{\alpha\rightarrow b}
    -\langle L^R_{\alpha\rightarrow b}\rangle)\\
    =Q^\infty_{\mathrm{fus}\rightarrow b}-Q_{b\rightarrow \mathrm{loss}}.
\end{align}

The total fusion power thus also splits into a gain and a loss term

\begin{align}
    \sum_{\ell}Q^\infty_{\mathrm{fus}\rightarrow \ell}-\sum_{\ell}Q_{\ell\rightarrow \mathrm{loss}}
    = Q_{\mathrm{fus}} - Q_{\mathrm{fus}\rightarrow \mathrm{loss}}.
\end{align}

\subsection{Neutron scattering}\label{sec:neutron-scattering}
The neutron energy deposition is calculated with an analogous model as
the alpha energy deposition. The only difference is that the formulas for
$\bold{R}_{b\ell}$ and $Q_{b\ell}$ in \eqref{euler2} and
\eqref{euler3} need to be changed. For the neutron ion interaction we
model the collisions as hard sphere collisions, as is implicitly
assumed in \cite{atzeni2004physics} and \cite{gus2011effect}. This
model slightly overestimates the neutron stopping since the the
differential cross sections for elastic collisions with ions are not
constant, as they would be for hard spheres, but smaller for high
scattering angles. Large scattering angles, however, result in high
energy transfer to the collision partners and a larger reduction of
the mean velocity. This is partially compensated by the fact that we
neglect inelastic scattering and also that we assume energy
independent cross sections and choose the value at $14 \mathrm{MeV}$,
although the cross section is considerably higher for smaller
energies. For hard spheres, the friction functions \(R, Q\) haven been worked
out in \cite[Eq. 15.14 and 15.15]{Burgers1969}
\begin{align}
  \label{Rbc_neutron}
    \bold{R}_{bc}&= -n_b m_b \nu_{bc} u_{bc} \Psi \left( \epsilon_{bc}
                   \right) \, , \\
  \label{Qcb_neutron}
  Q_{bc}&= \frac{m_b}{m_b+m_c} \, n_b \nu_{bc} \\
  &\hspace{0.4cm} \times \left[ 3 \left( kT_c- kT_b \right) \, \Phi \left(
    \epsilon_{bc} \right) + m_c u^2_{bc} \,
    \Psi \left( \epsilon_{bc} \right) \right] \, , \nonumber
\end{align}
where
\begin{align}
  \label{nubc_neutron}
    \nu_{bc}&= \frac{8}{3\sqrt\pi} \frac{n_c m_c}{m_b +
              m_c}  \sqrt{2 \left( \frac{kT_b}{m_b} + \frac{kT_c}{m_c}
              \right)} \, \sigma_{bc}
\end{align}
and
\begin{align}
  \label{Psi_neutron}
\Psi \left( \epsilon_{bc} \right)&=\frac{3\sqrt \pi}{8} \, \left( \epsilon_{bc} +
                \frac{1}{\epsilon_{bc}} - \frac{1}{4 \epsilon^3_{bc}} \right) \,
                                   \mathrm{erf} \left( \epsilon_{bc} \right) \\
  &\hspace{1cm} +\frac{3}{8}
                \left( 1+\frac{1}{2\epsilon^2_{bc}} \right)
                e^{-\epsilon^2_{bc}} \, , \nonumber \\
  \label{Phi_neutron}
\Phi \left( \epsilon_{bc} \right) &= \frac{\sqrt\pi}{2} \left( \epsilon_{bc} +
                 \frac{1}{2\epsilon_{bc}} \right) \mathrm{erf} \left(
                 \epsilon_{bc} \right) + \frac{1}{2}
                                    e^{-\epsilon^2_{bc}} \, ,
\end{align}
with
\begin{align}
  \label{epsilon_neutron}
\epsilon_{bc} &= \frac{|u_{bc}|}{\sqrt{2 \left( \frac{kT_b}{m_b} +
           \frac{kT_c}{m_c} \right)}} \, .
\end{align}
Here, \(u_{bc}=u_n\) since the ion species are considered to have no mean velocity in our model.
Making use of (\ref{euler2}) and (\ref{Rbc_neutron}) - (\ref{epsilon_neutron})
we obtain a coupled system of ODEs for the kinetic energy of a group of
neutrons with small angular spread in velocities and temperature analogous to the one for $\alpha$-stopping:
\begin{align}
  \label{stopping_single_neutron}
  \frac{d E_n}{dx}
  &= - \sqrt{2 m_nE_n} \, \sum_{c} \nu_{nc} \, \Psi \left(
    \epsilon_{nc} \right) \, , \\
  \label{temperature_single_neutron}
  \frac{d \, kT_n}{dx}
  &= \frac{2}{3} \, \sqrt{\frac{m_n}{2 E_n}} \, \sum_{c} \frac{m_n}{m_n+m_c} \, \nu_{nc} \\
  &\hspace{0.4cm} \times \left[ 3 \left( kT_c- kT_n
  \right) \Phi \left( \epsilon_{nc} \right) + \frac{2 \, m_c}{m_n} \, E_n
    \, \Psi \left( \epsilon_{nc} \right) \right] \, , \nonumber
\end{align}
where
\begin{align}
  \epsilon_{nc}
  &= \sqrt{\frac{m_c \, E_n}{m_n \, kT_c + m_c \, kT_n}} \, .
\end{align}
It is also possible to obtain the energy a typical neutron transfers to
the ion fluid $b$. We define
\begin{align}
\label{transferred_neutron_energy}
\zeta_b&= \frac{3}{2} \, \frac{n_b}{n_n} \, kT_b \, .
\end{align}
Replacing the index $n$  by $b$ and the index $c$ by $n$ in
(\ref{temperature_single_neutron}) and making use of
(\ref{transferred_neutron_energy}) we obtain
\begin{align}
  \label{zeta_eqns}
  \frac{d \zeta_b}{dx}
  &= \sqrt{\frac{m_n}{2E_n}} \, \frac{m_b}{m_n+m_b} \, \frac{n_b}{n_n}
    \, \nu_{bn} \\
  & \hspace{0.4cm} \times \left[ 3 \left( kT_n-kT_b \right) \, \Phi \left( \epsilon_{nb}
    \right) +  2 E_n \, \Psi \left( \epsilon_{nb} \right) \right] \, . \nonumber
\end{align}
Equations (\ref{zeta_eqns}) have to be solved in line with
(\ref{stopping_single_neutron}) and (\ref{temperature_single_neutron}).
The solution of these equations determines the heat transport per
neutron to species $b$. The average neutron energy deposition into species $b$ is obtained by 
substituting \(\frac{d\zeta_b}{dx}\), given by \eqref{zeta_eqns}, for \(\frac{d E_b}{dx}\) in \eqref{loss:alpha}. The result is  
\(\langle E^R_{n\rightarrow b}\rangle\). 

In the following, we approximate the above expressions for large velocities \(u_n\) in order to compare our model with elementary considerations. We obtain:
\begin{align}
  \label{Psi_approx}
  \Psi \left( \epsilon_{nc} \right)
  &=\frac{3\sqrt \pi}{8} \, \epsilon_{nc} \, , \quad
  \Phi \left( \epsilon_{nc} \right)
  = \frac{\sqrt\pi}{2} \epsilon_{nc} \, .
\end{align}
Treating all ions in the fuel as heavier than the
neutrons and that all scattering cross sections between neutrons and
fuel ions as approximately equal, i.e., $m_c \gg m_n$, $\sigma_{nc}
\approx \sigma_n$ and
\begin{align}
\label{Psi_nu_approx}
\sum_{c} \nu_{nc} \, \Psi \left( \epsilon_{nc} \right)
&\approx  \left| u_n \right| \, \sigma_{n} \, \sum_c n_c \, .
\end{align}
Making use of (\ref{stopping_single_neutron}), (\ref{zeta_eqns}),  and
(\ref{Psi_nu_approx}) leads to
\begin{align}
  \frac{dE_n}{dx}
  &\approx -\sqrt{2m_n E_n} \, \left| u_n \right| \, \sigma_{n} \, 
  \sum_c n_c  
\end{align}
and
\begin{align}
  \label{En_approx}
  \frac{dE_n}{dt}
  &\approx - 2 \, \left( \left| u_n \right| \, \sigma_{n} \, 
    \sum_c n_c \right) \, E_n \, , \\
  \label{zetab_approx}
  \frac{d \zeta_b}{dt}
  &\approx + 2 \,  \left( \frac{n_b}{n_n} \, \left| u_n \right| \, \sigma_{n} \, 
    n_b \right) \, E_n \, .      
\end{align}
Equations (\ref{En_approx}) and (\ref{zetab_approx}) can also be inferred
from elementary considerations. They imply that neutronic energy deposition
in non-cryogenic DT can be considerable and that neutrons heat light fuel
ions directly.

\subsection{Stationary temperature analysis}\label{sec:equilibrium}

Here we determine the quasi-equilibrium mentioned in sec. \ref{sec:ignition} by keeping the electron temperature fixed and studying the equilibrium temperatures that establish themselves for the ions via Eqs. \eqref{eq:temperaturedynamicsa}, \eqref{eq:temperaturedynamicsb}.
The validity of this quasi-equilibrium depends on the time scales of equilibration for ions and electrons. The flow diagram \ref{fig:MoreauBeBDTrhoR1} shows that the ions do, indeed, reach their equilibrium faster than the electrons. Note that in any case, the fixed points lie on the line of quasi-equilibrium, as at those points the electron temperature is constant. One can thus regard this notion of quasi-equilibrium as a simplified tool in the analysis of the temperature flow diagram \ref{fig:MoreauBeBDTrhoR1}.


The only relevant gain and loss mechanisms
for ions are the stopping of fusion alpha particles and neutrons
\eqref{fusionpower} as well as heat transfer to other particle species
\eqref{eq:Q_bc}.
Now, quasi-equilibrium is reached when gains and losses
exactly compensate each other. Hence, we have for all ion species \(i\):
\begin{align}\label{equilibrium}
    \sum_\ell Q_{i\ell}+Q_{\mathrm{fus}\rightarrow i}=0\,.
\end{align}

For \(n\) different species, one of them electrons with a given
temperature $kT_e$, \eqref{equilibrium} is a system of \(n-1\)
equations for \(n-1\) ion temperatures $kT_i$. This system of
equations can be solved numerically, yielding $n-1$ functions
$kT_i(kT_e)$.

In equilibrium, the ions receive energy from alpha particles and can
only pass on this energy to electrons. However, this is only possible
for \(kT_i>kT_e\). How much higher the different $kT_i$ are than $kT_e$
strongly depends on alpha and neutron stopping of the ion species. This effect
therefore varies with fuel composition and electron temperature. The
resulting difference in ion and electron temperatures is a useful
effect for fusion applications since it leads to an increased fusion
rate compared to radiation losses.

\begin{figure}[ht]
\begin{center}
  \includegraphics[width=.45\textwidth]{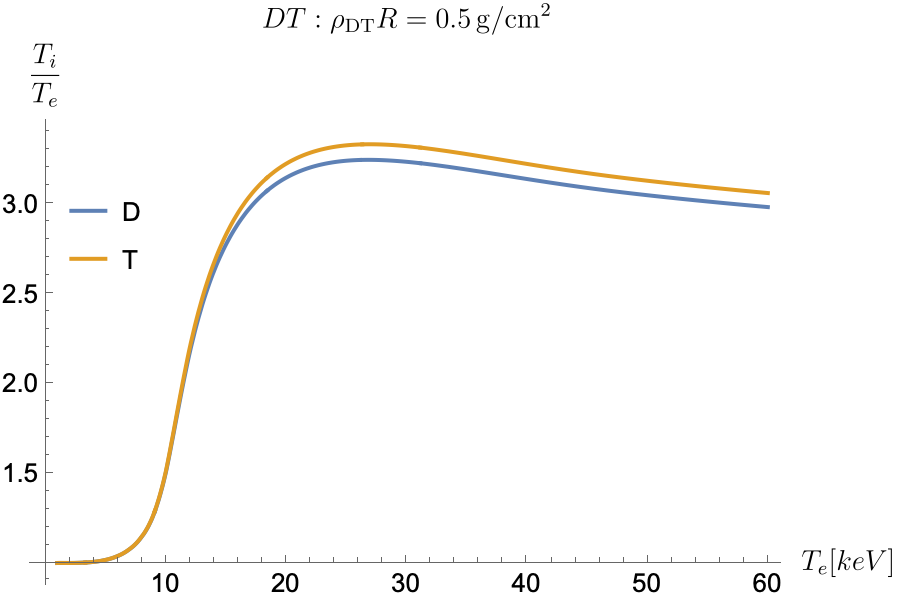}
\end{center}
\caption{\label{fig:DTTemperatureFrac} Ratio of 
  ion temperatures to electron temperature given by the solution of 
  \eqref{equilibrium} for a range of electron temperatures for 
  DT showing that $kT_{\ce{D}}, kT_{\ce{T}} \gg kT_{\ce{e}}$ for $kT_{\ce{e}} > 10 \, \ce{keV}$.}
\end{figure}

\begin{figure}[ht]
\begin{center}
  \includegraphics[width=.3\textwidth]{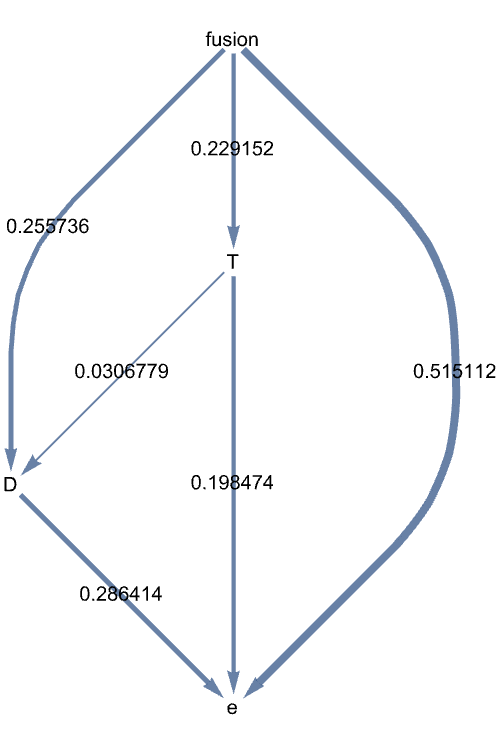}
\end{center}
\caption{\label{fig:DTGraph} Fraction of fusion power transferred 
  between the different species for an equal mixture of \(D\) and 
  \(T\) at \(\rho_{\mathrm{DT}} R=0.5g\mathrm{cm}^{-2}\). The ion temperatures in conditional equilibrium for $kT_e =
  30 \, keV$ are $kT_D=97.1\,keV$ and $kT_T=99.7\,keV$.}
\end{figure}

We illustrate the dependence of $kT_i$, $i=D,T$ on $kT_e$ in
Fig. \ref{fig:DTTemperatureFrac} for DT. At temperatures below
\(10\, \mathrm{keV}\), we see \(kT_i\approx kT_e\). Above this
threshold, the ionic temperatures rise up to \(kT_i\approx 3
kT_e\). The detailed balance of energy flux at \(kT_e=30\,
\mathrm{keV}\) is shown in Fig. \ref{fig:DTGraph}.

\subsection{Distribution of distances}\label{SEC:disofdis}
\begin{figure}
    \centering
    \includegraphics[width=0.5\textwidth]{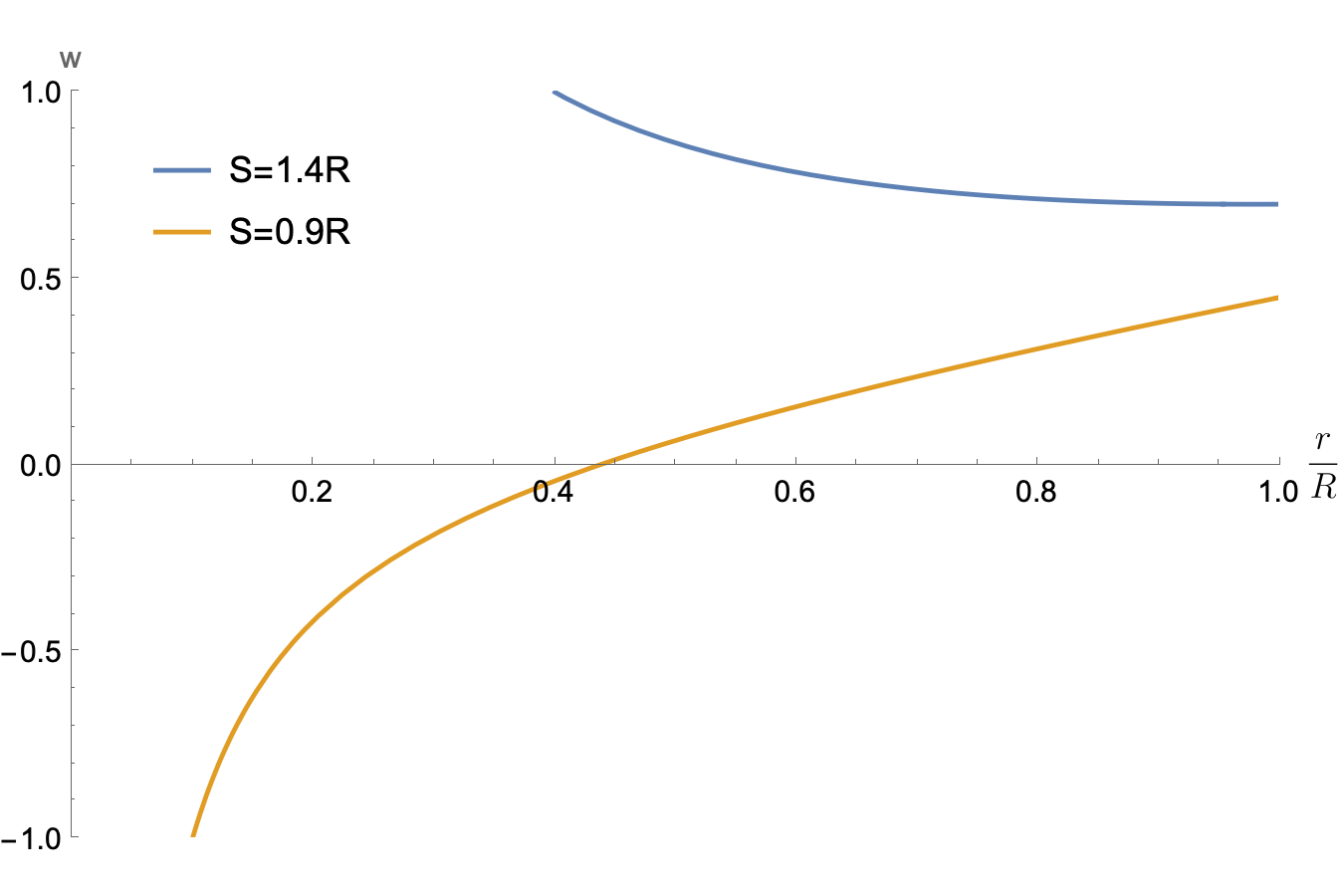}
    \caption{The integration boundary \eqref{separation:equation} for
      two different values of $S$. For $S=1.4R$ the area of
      integration $s\le 1.4R \wedge |w|<1\wedge 0<r<R $ is bounded
      from above by \(w=1\) for $r<0.4R$ and by
      \eqref{separation:equation} for $r>0.4R$. For $S=0.9R$
    the integration region $s\le 0.9R\wedge |w|<1\wedge 0<r<R$ begins
    at $r=0.1R$ and is bounded from above by
    \eqref{separation:equation}.}
    \label{fig:separationW}
\end{figure}

\begin{figure}
    \centering
    \includegraphics[width=0.5\textwidth]{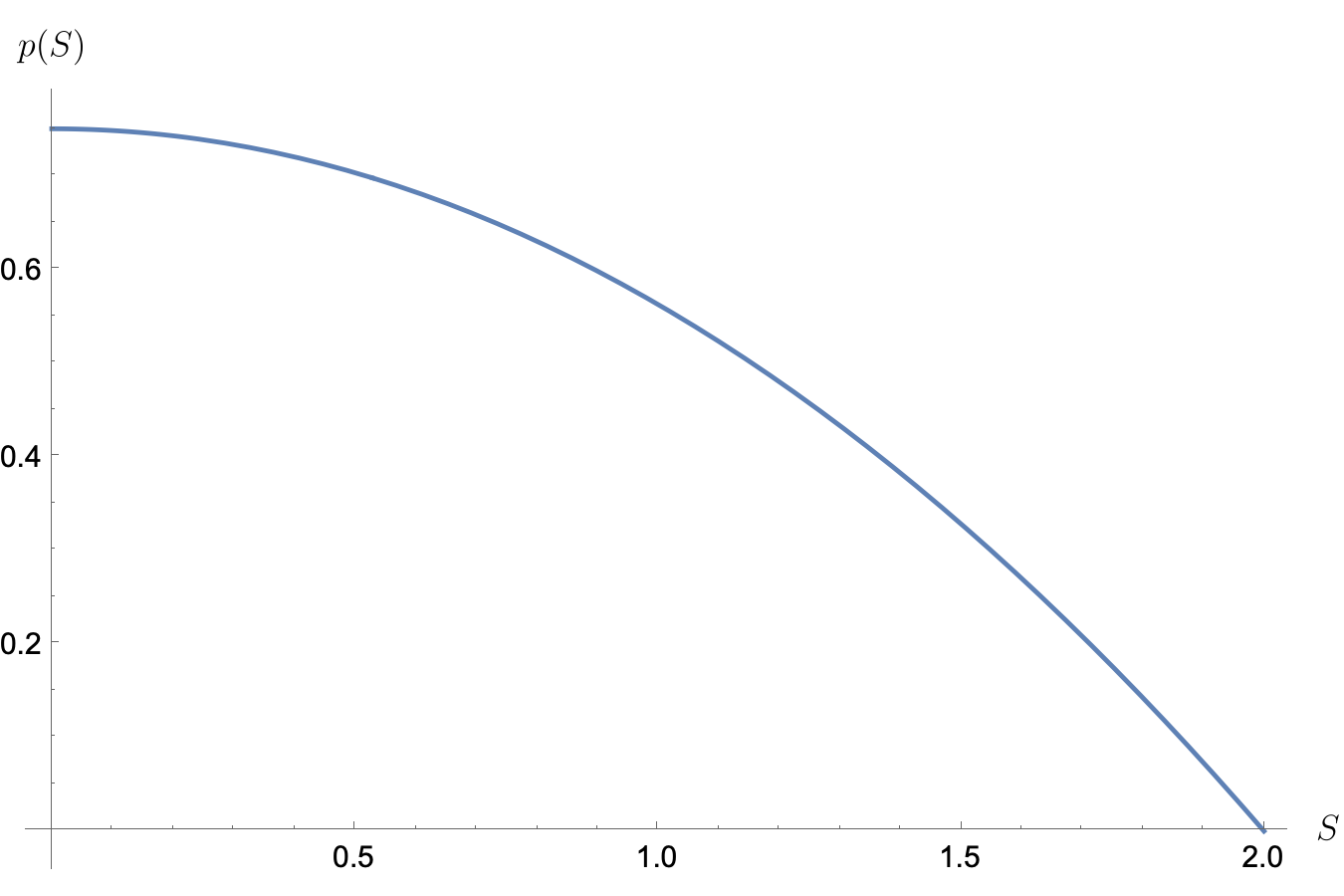}
    \caption{The probability distribution for the expected geometric
      flight distance $S$ of the alphas \eqref{sprob} for
      $R=1$. Without stopping power of the plasma, the alphas would on
      average cover a distance of $\frac{3}{4}R$ before leaving the
      fuel.}
    \label{fig:sprobalphas}
\end{figure}

Here we derive the distribution of geometric flight distances of alpha
particles inside a homogeneous ball of radius \(R\) similar to \cite{krokhin1973escape}. Hence, the
stopping of alpha particles will play no role in this derivation. We
assume that the alpha particles are equally likely to be created at
any point inside the ball, parameterised by the coordinates $r,
\theta_1$ and $\varphi_1$, and equally likely to have their initial
velocity point in any direction, parameterised by the coordinates
$\theta_2$ and $\varphi_2$. The angle $\theta_2$ for the flight
direction is chosen relative to the line connecting the initial
position of the particle and the center of the sphere.
The distance $s$ follows from the law of cosines
\begin{align}
    r^2+s^2-2rs\cos(\theta_2)=R^2\,.
\end{align}
We can write the cumulative distribution function of the distances as

\begin{align}
    P(s\le S) &= \frac{3}{(4 \pi)^2 R^3} \int_0^{2\pi}d\varphi_1 \int_0^\pi d\theta_1 \sin\theta_1 \int_0^R dr \, r^2 \\&
    ~~\int_0^\pi d\theta_2 \sin \theta_2 \int_0^{2\pi} d\varphi_2
  \mathds{1}_{s(r,\theta_2)\le S}\\
  &=\frac{3}{2 R^3} \int_0^R dr  \int_{-1}^1 dw ~ r^2 \mathds{1}_{s(r,w)\le S}.
\end{align}
The boundary between the regions $s(r,w)\le S$ and $s(r,w)>S$ in the
$r,w$ coordinate space is given by
\begin{align}\label{separation:equation}
    w=\frac{r^2+S^2-R^2}{2rS}\,.
\end{align}
For any given value of \(S\), the point \((r,w)=(R,1)\) in the upper
right corner of the integration domain corresponds to the maximal
value \(s(r,w)=2R\). Therefore, it is inside the region \(s(r,w)>S\)
that is not contributing to the integration for all values
\(S<2R\). Therefore \(w\) should be integrated over all values less
than the one singled out by Eq.\ \eqref{separation:equation}. The
boundary Eq.\ \eqref{separation:equation} intersects the boundaries of
the integration domain at $r=S-R, w=1$ for $S>R$ and at $r=R-S,w=-1$
for $S<R$ as exemplified in Fig. \ref{fig:separationW}. This leads to
the two cases
\begin{align}\label{cases}
    P(s\le S)=\frac{3}{2R^3}\left\{\begin{matrix*}[l]
        \int_0^{S-R}dr r^2\int^1_{-1}dw&S>R\\
        \hspace{0.5cm}+\int^R_{S-R}drr^2\int_{-1}^\frac{r^2+S^2-R^2}{2rS} dw  &\\
        \int_{R-S}^R dr r^2 \int_{-1}^\frac{r^2+S^2-R^2}{2rS} dw & S<R
    \end{matrix*}\right.
\end{align}
Both cases lead to the same value 
\begin{align}
    P(s\le S)=\frac{S(12 R^2 - S^2)}{16 R^3}
\end{align}
resulting in the probability density
\begin{align}\label{sprob}
    p(S)=\frac{\partial}{\partial S} P(s\le S)=\frac3{4R} - \frac3{16} \frac{S^2}{R^3}\,,
\end{align}
which is shown in Fig. \ref{fig:sprobalphas}.

\subsection{Reactivity for mixed temperatures} \label{sec:mixedtempreactivity}
In this section, we describe how to obtain the reactivity for a fusion
reaction between two different ion species, $i$ and $\ell$, which have
different temperatures $kT_i$ and $kT_\ell$. This is necessary as we allow for different temperatures of the different ionic species.

One needs to calculate
\begin{align} \label{eq:sigmav}
    \langle \sigma(v_{rel}) v_{rel} \rangle=\int f_i f_\ell \sigma(v_{rel}) v_{rel} d^3v_i d^3v_\ell
\end{align}
where
\begin{align}
    f_b=\frac{n_b}{\left(2\pi kT_b/m_b\right)^{3/2}}\exp\left(-\frac{m_bv_b^2}{2kT_b}\right)
\end{align}
for $b = i, \ell$. To simplify \eqref{eq:sigmav}, the center of mass and relative
velocity coordinates need to be modified and temperature weighted,
leading to
\begin{align}
  \mathbf{U}&=\frac{\frac{m_i}{kT_i}\mathbf{v}_i+\frac{m_\ell}{kT_\ell}\mathbf{v}_\ell}{\frac{m_i}{kT_i}+\frac{m_\ell}{kT_\ell}},\\
    \mathbf{v}_{rel}&=\mathbf{v}_i-\mathbf{v}_\ell,\\
    f_i f_\ell&=\frac{n_i n_\ell}{\left(4\pi^2  kT_i kT_\ell/(m_i m_\ell)\right)^{3/2}}\nonumber\\ &\times \exp\left(-\left(\frac{m_i}{kT_i}+\frac{m_\ell}{kT_\ell}\right)\frac{U^2}{2}
    -\frac{\frac{m_i}{kT_i}\frac{m_\ell}{kT_\ell}}{\frac{m_i}{kT_i}+\frac{m_\ell}{kT_\ell}} \frac{v_{rel}^2}{2}\right)\,.\label{tempweightcoor}
\end{align}
For equal temperatures, the prefactor of $v_{rel}^2$ in
\eqref{tempweightcoor} is given by $\mu/(2kT)$, where $\mu = m_i
m_\ell/(m_i+m_\ell)$ is the reduced mass. A comparison with
\eqref{tempweightcoor} shows that the reactivity in the two
temperature case can be reduced to that for a single temperature
given by
\begin{align}
    \label{mixtemp}
    kT=\frac{m_i kT_\ell+m_\ell kT_i}{m_i+m_\ell}\,.
\end{align}

\subsection{Radiation Losses}
Based on a similar "ray tracing" model as for the alpha particles, we here determine how much of the radiation gets reabsorbed due to the opacity of the fuel. Moreover, we compare the result to another analytical model based on the radiation diffusion equations that MULTI implements for a simplified case.

\begin{figure*}
    \centering
    \includegraphics[width=0.8\textwidth]{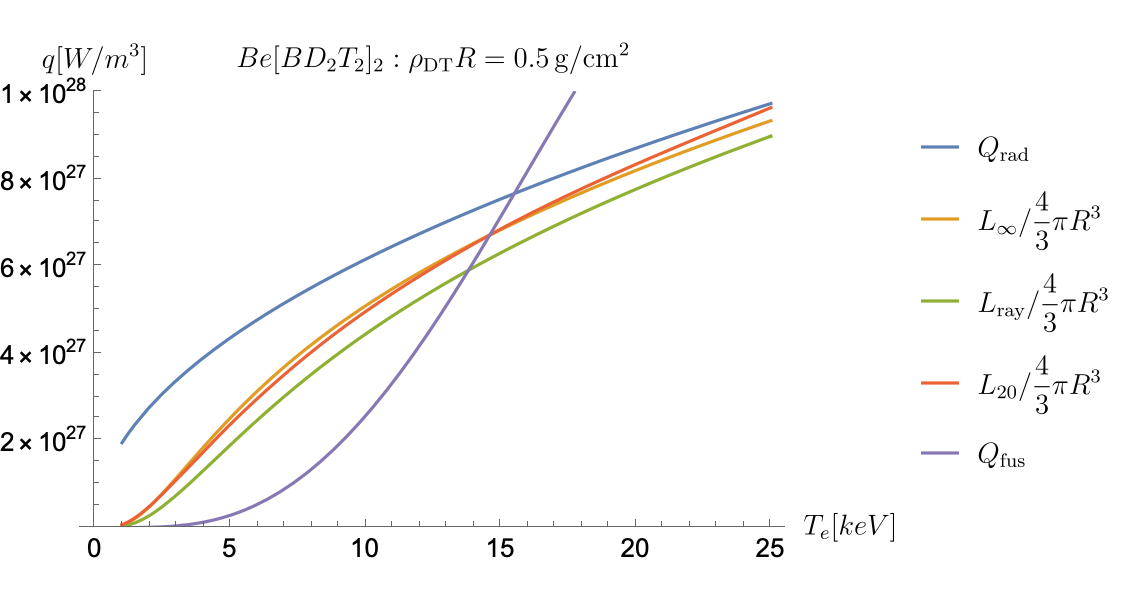}
    \caption{Total radiation power density according to the different
      models together with the fusion power density for
      $\ce{Be[BD_2T_2]_2}$, $\rho_\mathrm{DT}=50 \, \ce{gcm^{-3}}$,
      and $R=0.01 \, \ce{cm}$. The unmitigated radiation loss
      power $Q_{\ce{rad}}$, which is used in the main text leads to
      the highest radiation losses. The effect of radiation
      re-absorption is strongest at low temperatures as the
      $T^{-3.5}$ dependence of the opacity \eqref{opacity}
      suggests. The different results $L_{\ce{ray}}$
      \eqref{eq:lray} and the multi group results $L_{20}$ and
      $L_\infty$ \eqref{eq:lg} agree closely. For the fusion power,
      the quasi equilibrium temperatures have been used.}
    \label{fig:radLoss}
\end{figure*}

We are interested in the radiation power for the frequency $\nu$ that
escapes the fuel sphere. To this end, we imagine that each photon is
generated at a randomly uniform position within the fuel sphere and
with a randomly uniform flight direction. Since this is analogous our
previous model for alpha particles and neutrons, we can re-use formula
\eqref{sprob} for the distribution of flight distances. Then $L_\nu$
can be obtained by integrating the emissivity $\eta_\nu$ over all
initial positions (factor $\frac{4\pi}{3}$), flight directions (factor
$4\pi$) and flight distances $s$ weighted with the
probability \eqref{sprob} for the flight distance $s$ multiplied by
the probability $\exp(-\kappa_\nu's)$ that the photon does not get
absorbed after a distance $s$. Here,
$\kappa'_\nu=(1-\exp(-h\nu/kT))\kappa_\nu$ is the relevant opacity
(compare \cite[chap. 7.3.2]{atzeni2004physics}) which is related to
$\eta_\nu$ via Kirchhoff's law $\eta_\nu=\kappa'_\nu I_{P\nu}$ where
$I_{P\nu}$ is the Planck intensity. This yields
\begin{align}
  L_\nu
  &=\frac{4}{3} \pi R^3 \eta_\nu \, 4\pi \int_0^{2R}ds 
  \left(\frac3{4R} - \frac3{16} \frac{s^2}{R^3}\right)
  \exp(-\kappa'_\nu s) \\
 &=4\pi^2R^2I_{P\nu}\left(1-\frac{1-\exp(-2\kappa'_\nu
 R)(1+2\kappa'_\nu R)}{2{\kappa'}^2_\nu R^2}\right) \, . \nonumber
\end{align}
With the expression for the opacity \cite[eqs. 10.91,
10.92]{atzeni2004physics} corresponding to \eqref{loss:rad},
\begin{align}
  \label{opacity}
  \kappa'_\nu
  &=\left[1-\exp\left(-h\nu/(kT)\right)\right] \, \\
  &\times 
  \frac{2\sqrt{3}}{\pi}\frac{16\pi}{3\sqrt{6\pi}}\frac{e^6}{\left(4\pi \epsilon_0\right)^3 m_e^{3/2}c}\frac{n_e
  \sum_i Z_i^2 n_i}{\sqrt{kT}}\frac{1}{2h\nu^3} \nonumber \\
  &=\underbrace{\frac{16}{3\sqrt{2\pi}}\frac{e^6}{\left(4\pi \epsilon_0\right)^3 m_e^{3/2}c}
  \frac{n_e \sum_i Z_i^2 n_i}{(kT)^{7/2}}h^2R}_{=:x}
  \frac{1}{R}\frac{1-\exp(-u)}{u^3} \nonumber \\
  &=\frac{x}{R} \frac{1-\exp(-u)}{u^3} \nonumber
\end{align}
it follows that the total radiation loss power estimated by "ray tracing" is given by
\begin{align}
  &L_\mathrm {ray}
   :=\int_0^\infty d\nu \,  L_\nu 
   = 4\pi^2R^2\left(\frac{kT}{h}\right)^4\frac{2h}{c^2} \int_0^\infty
  du \, \frac{u^3}{\exp(u)-1} \nonumber \\
  &\times \left(1-\frac{1-\exp\Big(-2x\frac{1-\exp(-u)}{u^3}\Big)\Big(1+2x\frac{1-\exp(-u)}{u^3}\Big)}
  {2\left(x\frac{1-\exp(-u)}{u^3}\right)^2}\right) \, .
  \label{eq:lray}
\end{align}
This integral is approximately equal to
$\text{min}(\frac{4}{3}\log(1+x),\frac{\pi^4}{15})$, but we use
numerical solutions. We estimate
\begin{align}\label{thin}
  \text{for }x \ll \text{1:     } L_\mathrm {ray}
  &=\frac{4}{3}\pi R^3Q_\mathrm{rad} \, , \\
  \label{thick}
  \text{for }x \gg \text{1:     } L_\mathrm {ray}
  &=4\pi R^2 \sigma T^4 \, ,
\end{align}
where $Q_{rad}$ is given by Eq. \eqref{loss:rad} and $\sigma$ is the
Stefan-Boltzmann constant. Hence, \eqref{eq:lray} interpolates between
optically thin plasma for $x\ll 1 $ and optically thick
plasma for $x\gg 1 $. It can be seen in
Fig. \ref{fig:radLoss} that especially at lower temperatures the
radiation re-absorption does reduce the radiation losses
considerably. However, we emphasize that the results for the
analytical model in the main text do not take into account radiation
re-absorption as it breaks the simple $\rho_{\ce{DT}}R$
scaling\footnote{This can be seen from the fact that $x$ in
  \eqref{opacity} depends on $\rho^2R$.}. We hence obtain a
conservative estimate.

The MULTI-IFE code, however, uses a different method to estimate the
radiation losses. It solves the multi-group radiation equations ($i$
being the radiation group index)
\begin{align}
    \nabla\cdot\bold{S}^i&=c\kappa^{Pi}(U^{Pi}(T)-U^i) \, , \\
    \label{diff}
    \bold{S}^i&=-\frac{c}{3\kappa^{Ri}}\nabla U^i \, .
\end{align}
For the simple scenario of a hot spherical plasma with constant
temperature $T$ and constant Rosseland and Planck opacities
$\kappa_h^{Ri},\kappa_h^{Pi}$ surrounded by a cold medium with $T=0$
and also constant opacities $\kappa_c^{Ri},\kappa_c^{Pi}$, those
equations can be solved analytically
\begin{eqnarray}
  \Delta U^i
  &=&
    \frac{1}{r}\partial_r^2(rU^i)=3\kappa^{Ri}\kappa^{Pi}(U^i-U^{Pi}(T))\\
  U_h^i
  &=&
      c^i_1\frac{\exp(\sqrt{3\kappa^{Ri}_h\kappa^{Pi}_h}r)}{r} \\
  &&+c^i_2
     \frac{\exp(-\sqrt{3\kappa^{Ri}_h\kappa^{Pi}_h}r)}{r}+U^{Pi}(T)
     \nonumber \\
  U^i_c
  &=&
      c^i_3\frac{\exp(\sqrt{3\kappa^{Ri}_c\kappa^{Pi}_c}r)}{r} \\
  &&+c_4^i
  \frac{\exp(-\sqrt{3\kappa^{Ri}_c\kappa^{Pi}_c}r)}{r} \, . \nonumber
\end{eqnarray}
Here, $U^{Pi}(T) = \frac{4\pi}{c} \int_{\nu_{i-1}}^{\nu_i} \, I_{P\nu}(T)$.
Physical solutions require $c^i_3=0$ and that $c^i_2=-c^i_1$ and
continuity of $U^i$ and $\bold{S}^i$ at $r=R$, from which
$c^i_1$ and $c^i_4$ can be determined. With \eqref{diff}, the total
radiation outflow $\bold{S}^i(R)$ of the hot sphere is
determined. Here, the case of a surrounding vacuum is needed. It is
obtained by taking the limit $\kappa_c^{Ri},\kappa_c^{Pi} \rightarrow 0$
for $\bold{S}^i(R)$, leading to
\begin{align}
  \bold{S}^i(R)
  &=\left(\sqrt{\frac{\kappa^{Pi}_h}{3\kappa^{Ri}_h}}
  \coth(\sqrt{3\kappa^{Ri}_h\kappa^{Pi}_h}R)-\frac{1}{3\kappa^{Ri}_h
    R}\right)cU^{Pi}(T) \, .
\end{align}
In the limit of infinitely many groups, we have
$\kappa^{Ri}_h=\kappa^{Pi}_h=\kappa'_v$. Using the analytic expression
\eqref{opacity} for $\kappa'_v$, we obtain
\begin{align}
  L_\infty &=\int_0^\infty d\nu \, 4\pi R^2 \, \bold{S}^i(R)\\
  &=R^2\left(\frac{kT}{h}\right)^4\frac{32\pi^2 h}{c^2}\int_0^\infty
    du\frac{u^3}{\exp(u)-1} \nonumber \\
  &\times
    \left(\frac{1}{\sqrt{3}}\coth\left(\sqrt{3}x\frac{1-\exp(-u)}{u^3}\right)
     -\frac{u^3}{3x(1-\exp(-u))}\right) \,. \nonumber
\end{align}
While this formula has the same optically thin limit \eqref{thin}, in
the case for optically thick plasma \eqref{thick}, there is an
additional factor $4/\sqrt{3}$. Since this case is not relevant for the scenarios investigated here, 
we abstain from further discussing this factor.

For a finite number $G$ of groups the total radiation loss power is
given by
\begin{align}
\label{eq:lg}
  L_\mathrm {G}
  = \, &4\pi R^2 \sum_i^\mathrm
    {G}\left(\sqrt{\frac{\kappa^{Pi}_h}{3\kappa^{Ri}_h}}
      \coth(\sqrt{3\kappa^{Ri}_h\kappa^{Pi}_h}R) -\frac{1}{3\kappa^{Ri}_h R} \right)\\
  &\times
  \int_{\nu_{i-1}}^{\nu_i}d\nu\frac{8\pi h
    \nu^3}{c^2}\frac{1}{\exp(h\nu/kT)-1} \, . \nonumber
\end{align}
To compare more directly with the MULTI code, we evaluate this
analytically determined function using the same numerical opacity
tables
\cite{magee1995atomic}(\hyperlink{https://aphysics2.lanl.gov/apps/index.py}{TOPS
  Opacities}). Fig. \ref{fig:radLoss} shows that the different models
for the radiation losses are in close agreement.

\section{Modifications to MULTI-IFE}\label{sec:multi1D_modified}
\begin{figure}
    \centering 
    \includegraphics[width=0.5\textwidth]{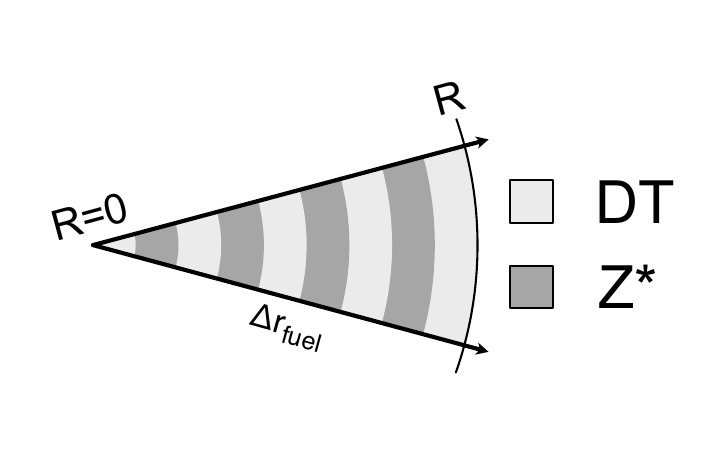}
    \caption{Alternating layers setup. The simulation volume
    $r<r_\mathrm{fuel}$ is filled with alternating layers of reactive
    (DT) and non-reactive (Z*) materials.}
    \label{fig:simulation-grid}
  \end{figure}

To accommodate active and inactive mass densities $\rho_{\ce{DT}}$
and $\rho_{\ce{Z}}$ within the framework of MULTI-IFE without substantial
modifications to the code at this time we introduce alternating layers of active
$\ce{DT}$ and inactive $\ce{Z}$ materials (see Fig. \ref{fig:simulation-grid}).

The layering changes the balance of fusion power and radiation losses
compared to a homogeneous mixture with a certain value of
$\rho_{\ce{DT}} R$. To create a layered scenario which is comparable
with the homogeneous mixture, we use DT layers with the density
$2\rho_{\ce{DT}}$ and Z* layers with the density $2\rho_{\ce{Z}}$ as
well as the radius of the layered construction $R_{\ce{layers}}$ being
equal to the unlayered one. Then the average densities are the same in
both the layered and unlayered cases.

We consider a layer of thickness $2 \Delta R$ of the homogeneous
mixture. In the layered construction it gets replaced by a DT layer
and a Z* layer, each of thickness $\Delta R$. The fusion power
generated by this layered construction is increased by a factor of
two compared to the homogeneous mixture in the unlayered construction
as the DT density is doubled (and enters fusion power quadratically)
but occupies only half the volume \cite{karlchenvomIPP6}.

However, through the layering the radiation rate is increased
too by a factor which depends on the fuel mixture. For the case
of $\ce{Be[BD_2T_2]_2}$ we have
\begin{align}
  \frac{Q_{\mathrm{rad,layers}}}{Q_{\mathrm{rad,hom}}}
  &= \frac{\left(2n_e \sum_i Z_i^2 
    2n_i\right)_{\mathrm{D_4T_4}}+\left(2n_e \sum_i Z_i^2 
    2n_i\right)_{\mathrm{BeB_2}}}{2\left(n_e \sum_i Z_i^2 
    n_i\right)_\mathrm{BeB_2D_4T_4}} \\
   &=\frac{494}{407} \approx 1.2 \, , \nonumber 
\end{align}
which is smaller than the increase in the fusion rate. 
To obtain the correct balance between fusion and radiation 
powers we have thus chosen to increase the opacities by the
quotient of these two factors, i.e., by $2/1.2 \approx 1.65$.
By this approach the radiation losses are effectively increased
compared to the fusion power. We note that other power terms
may also get modified by the layering process. However, fusion
power and radiation loss are the dominant terms controlling
the ignition of the fuel. We note that due to the linearity of the
stopping power $\frac{dE}{dx}$ in the densities, the layered scenario
produces the same stopping behavior of alpha particles as the
homogeneous one. Moreover, also the input and output energies
are the same.

MULTI-IFE is a code designed for $\ce{DT}$. As our analytical analysis
shows (see Fig. \ref{fig:BeB_channels}), non-cryogenic DTs have
a non-negligible deposition power contribution from the stopping of $\alpha$-particles
in the ionic background that increases with fuel temperature, leading
to a substantial direct fuel ion heating with growing fuel temperature.
At present, MULTI-IFE does not account for this effect.
In addition, fusion neutrons deposit non-negligible fractions
of their energy in non-cryogenic DTs. Moreover, neutronic energy
deposition in the fuel is not considered.
As a consequence, the MULTI-IFE simulations given in
Fig. \ref{fig:report_tite_a1} underestimate the fusion energy gain
obtainable from non-cryogenic DTs.

\end{document}